\documentclass[12pt]{article}
\usepackage{graphicx,amssymb,amsmath}
\usepackage{epsfig}
\usepackage{color,graphicx}
\usepackage{cite}
\usepackage{amssymb,amsmath}
\usepackage{multirow}
\usepackage[top=2.5cm, bottom=2.5cm, left=2.5cm, right=2.5cm]{geometry}	

\numberwithin{equation}{section}

\newcommand{\nn}{\nonumber}

\newcommand{\order}[1]{${\cal O}\left(#1 \right)$}
\newcommand{\morder}[1]{{\cal O}\left(#1 \right)}

\newcommand{\be}{\begin{equation}}
\newcommand{\ee}{\end{equation}}
\newcommand{\bea}{\begin{eqnarray}}
\newcommand{\eea}{\end{eqnarray}}

\begin{document}
 
\begin{flushright}
HIP-2021-39/TH\\
OW - 2021-1\\
APCTP Pre2021 - 030
\end{flushright}

\begin{center}

\centering{\Large {\bf Novel semi-circle law and Hall sliding \\
in a strongly interacting electron liquid}}

\vspace{8mm}

\renewcommand\thefootnote{\mbox{$\fnsymbol{footnote}$}}
Niko Jokela,${}^{1,2}$\footnote{niko.jokela@helsinki.fi}
Matti J\"arvinen,${}^{3,4}$\footnote{matti.jarvinen@apctp.org} and
Matthew Lippert${}^{5}$\footnote{lippertm@oldwestbury.edu}

\vspace{4mm}
${}^1${\small \sl Department of Physics} and ${}^2${\small \sl Helsinki Institute of Physics} \\
{\small \sl P.O.Box 64} \\
{\small \sl FIN-00014 University of Helsinki, Finland} 

\vspace{2mm}

${}^3${\small \sl  Asia Pacific Center for Theoretical Physics} \\
{\small \sl Pohang 37673, Republic of Korea}

\vspace{2mm}

${}^4${\small \sl Department of Physics} \\
{\small \sl Pohang University of Science and Technology} \\
{\small \sl Pohang 37673, Republic of Korea}

\vspace{2mm}

${}^5${\small \sl Department of Chemistry and Physics} \\
{\small \sl SUNY Old Westbury} \\
{\small \sl Old Westbury, New York, USA} 

\end{center}

\vspace{8mm}

\setcounter{footnote}{0}
\renewcommand\thefootnote{\mbox{\arabic{footnote}}}

\begin{abstract}
\noindent

We study a strongly interacting, fermionic fluid in the presence of an applied magnetic field using a holographic framework.  At low temperatures, translation symmetry is spontaneously broken and the resulting phase is a striped Hall fluid.  Due to the magnetic field, an electric field applied parallel to the stripes causes the stripes to slide, a phenomenon we coin ``Hall sliding.'' We also investigate the magneto-transport of the system in the presence of an explicit translation symmetry-breaking lattice which pins the stripes. Electrical properties are well represented by a hydrodynamical model, which gives us further insight into particle-like cyclotron and pseudo-Goldstone excitations we observe. The DC conductivities obey a novel semi-circle law, which we derive analytically in the translationally invariant ground state at low temperature.

\end{abstract}



\newpage

\section{Introduction}
\label{sec:introduction}

When subjected to a magnetic field, strongly correlated electrons exhibit a range of interesting and puzzling behaviors.  The fractional quantum Hall (FQH) effect, featuring gapped phases at non-integer filling fraction $\nu$ and with fractionally charged quasiparticles, is perhaps the most famous.  Although phenomenological descriptions of FQH states can be given in terms of Laughlin wavefunctions or fermions with attached flux quanta, a comprehensive theoretical description, particularly including transitions between gapped quantum Hall states, remains elusive.

Although many FQH states have been observed at small filling fractions below the first Landau level, they become increasingly sparse at filling fractions $\nu > 2$.  Near half-filling between high Landau levels ($\nu  = N + 1/2$ for $N > 4$), interactions can lead to a striped, charge density wave (CDW) ground state, which exhibits a strongly anisotropic charge transport \cite{PhysRevLett.82.394,DU1999389,von_Oppen_2001}.  

Under general assumptions and independent of microscopic details, the charge conductivity in between quantum Hall states obeys non-trivial relations, notably, the semi-circle law \cite{Burgess:1999ug}
\be
\label{eq:QHsemicircle}
\sigma_{xx}^{DC}\sigma_{yy}^{DC}\ + \left(\sigma_{yx}^{DC}\ - \sigma^0_{h} \right)^2 = \left(e^2/2h\right)^2
\ee
where $\sigma^0_{h} = \frac{e^2}{h} (N+1/2)$ and the end points of the semi-circle are quantum Hall states.  This robust relation holds away from symmetric half-filling point, in the presence of disorder, and in wide class of striped systems \cite{MacDonald_2000,von_Oppen_2000}. 

As a particular example of a strongly correlated electronic material, high $T_c$ cuprate superconductors demonstrate a number of unexplained features.  For example, the Hall angle $\theta_H = \tan^{-1}\left( \sigma_{xy}^{DC} / \sigma_{xx}^{DC}\right)$,  which measures the transverse Hall conductivity relative to the longitudinal Ohmic conductivity, exhibits surprising temperature scaling. Although the Hall angle might be expected to have the same temperature dependence as the longitudinal DC conductivity, which for cuprates scales inversely with temperature $\sigma^{DC} \sim T^{-1}$, 
the Hall angle instead is found experimentally to scale as $\tan \theta_{H} \sim T^{-2}$, which is typical of normal metals.

Gauge/gravity duality provides a tractable method of analyzing strongly interacting systems.  In particular, observables such as charge transport can be reliably computed, and there has been a strong focus on magneto-transport of holographic models in recent years.  An early holographic-inspired hydrodynamical model of magneto-transport was constructed in \cite{Hartnoll:2007ih}, followed by a number of other holographic models \cite{Pal:2010sx, Blake:2014yla, Amoretti:2015gna, Blake:2015ina, Amoretti:2016cad, Cremonini:2017qwq, Blauvelt:2017koq, Cremonini:2018kla, Hoyos:2019pyz, Amoretti:2019buu, Amoretti:2020mkp, Song:2019rnf, Baggioli:2020edn, Jeong:2021zhz, Jeong:2021zsv}, with attempts made to match the characteristic strange metallic behavior.

With the goal of investigating the striking magnetic phenomena of strongly correlated electrons in an explicit top-down holographic construction, we study the magneto-transport of the D3-D7' probe-brane model, which holographically describes a strongly coupled (2+1)-dimensional electron fluid.\footnote{Magnetic fields and the quantum Hall effect have also been investigated in a number of other top-down  \cite{Davis:2008nv, Fujita:2009kw, Alanen:2009cn, Jokela:2011eb, Kristjansen:2012ny, Bea:2014yda} and bottom-up holographic models \cite{Keski-Vakkuri:2008ffv, Bayntun:2010nx, Goldstein:2010aw, Lippert:2014jma, Dolan:2021gtj}.}  
The gapless phase of this model has been 
suggested as a holographic description for graphene \cite{Davis:2011gi,Omid:2012vy,Semenoff:2012xu} or other strongly interacting 2D systems, such as the recently proposed scandium Herbertsmithite \cite{DiSante:2019zrd}.  However, a direct application of this type of (2+1)-dimensional defect model to graphene would be problematic since the 3D Coulomb interaction is weakly coupled to the graphene and the speed of light is much higher than the Fermi velocity in graphene, indicating significantly different physics. 	

The D3-D7' system has a quantum critical point at zero temperature, magnetic field, and charge density, which indicates a phase transition between a homogeneous phase and a spatially modulated striped phase.  This striped phase exhibits spontaneous, intertwined striped order, including charge density waves and spatially modulated, transverse, persistent currents. Because the translation symmetry breaking is spontaneous, an applied electric field causes the stripes to slide, leading to collective charge transport \cite{Jokela:2016xuy}.  Somewhat surprisingly, the striped phase with this collective sliding mode was found to have only mildly different conductivities compared to the homogeneous phase.  However, the stripes can be pinned by the introduction of an explicit background lattice, which significantly impacts the transport across the stripes \cite{Jokela:2017ltu}.

The D3-D7' model exhibits many interesting properties in the presence of a background magnetic field.  At nonzero charge density, the system is in a gapless metallic state, but for certain values of the magnetic field, gapped quantum Hall states can appear \cite{Bergman:2010gm}.   At nonzero fermion mass, the gapless phase displays ferromagnetism and an anomalous Hall effect  \cite{Jokela:2012vn}.  The magnetic field also stabilizes the homogeneous phase against the instability to form stripes \cite{Jokela:2012vn, Jokela:2014dba}.

The charge conductivity was first studied in \cite{Bergman:2010gm }, where only the DC conductivities of the homogeneous gapped and ungapped phases were computed.  In \cite{Jokela:2012vn}, the high-temperature behavior of DC conductivity in the gapless phase was further analyzed.   In the related D2-D8' model, which similarly features a gapped, quantum Hall state at nonzero charge density, the DC conductivity of the homogeneous phases was shown to partially reproduce the semi-circle law observed in quantum Hall transitions \cite{Jokela:2011eb}.


In this paper, we extend this earlier analysis and fully analyze the conductivity of the D3-D7' model in an applied magnetic field, in both the gapless homogeneous and striped phases.  Furthermore, we also investigate the effects of pinning by adding an explicit symmetry-breaking lattice.  The optical conductivities are obtained numerically by solving for the linearized fluctuations of the numerically computed background. The DC conductivities, however, can be computed analytically in terms of the numerical horizon data of the background solutions.  The numerical AC conductivities are shown to match the semi-analytic DC conductivities in the zero-frequency limit.

The DC conductivity is found to obey a semi-circle law, but not the usual law \eqref{eq:QHsemicircle} observed in quantum Hall transitions.  For the D3-D7' model, the roles of $\sigma_{xx}^{DC}$ and $\sigma_{xy}^{DC}$ are reversed compared with \eqref{eq:QHsemicircle}, and, unlike \eqref{eq:QHsemicircle}, the radius of the semi-circle is proportional to the charge density over temperature squared.  In the low-temperature limit, we are able to derive this semi-circle law analytically for the homogeneous phase:
\be\label{eq:oursemicirclelaw}
   \left(\sigma_{yy}^{DC}-\frac{\sqrt{2}\pi D}{\sqrt\lambda  N_c T^2}\right)^2 + \left(\sigma_{xy}^{DC}\right)^2 = \left(\frac{\sqrt{2}\pi D}{\sqrt\lambda  N_c T^2}\right)^2 \ , 
\ee
where $D$ is the physical charge density and the division by the number of colors $N_c$, $D/N_c$, can be viewed as charge density per fermion species. The 't Hooft coupling $\lambda$ can be written in terms of field theory data $\lambda=g_{YM}^2N_c$ of the ambient ${\cal N}=4$ super Yang-Mills theory. Numerical evidence indicates the behavior continues to good approximation away from the low-temperature limit.  Note that in here, this novel semi-circle law holds exactly in the limit of large charge and finite magnetic field, that is, at large filling fraction, while the usual semi-circle law is found at order-one filling fraction.

For the striped phase, the semi-circle law persists, albeit for the spatially averaged conductivities and with a modified radius.  Because the striped phase is anisotropic, there is an ambiguity in the direction chosen for the longitudinal conductivity.  However, the DC conductivities across and along the stripes are approximately equal, so different choices result in a small change to the semi-circle radius.  When the stripes are pinned, the conductivities are strongly anisotropic but still obey an exact relation which can be interpreted as an anisotropic analog of the semi-circle law. 

The Hall angle $\theta_H$ obtained from the DC conductivities shows the scaling $\tan\theta_H\sim T^{-2}$ at low temperatures.  However, the longitudinal conductivity $\sigma_{xx}^{DC}$ does not exhibit the characteristic $T^{-1}$ scaling of strange metals.

Previously, we found \cite{Jokela:2016xuy} that the stripes of the inhomogeneous phase slide due to an applied electric field.  With addition of a magnetic field, the stripes now slide not just when an electric field is applied across the stripes but also when it applied along them.  We find that this ``Hall sliding'' has a different velocity than the standard sliding studied in \cite{Jokela:2016xuy}.  While the velocity of the usual sliding increases as the magnetic field grows, the velocity of the Hall sliding decreases.


In addition to DC conductivities, we also investigate the optical, or AC, conductivities, which give us access to physics associated with nonzero frequency.  Our results can be fit well by the hydrodynamical model of \cite{Hartnoll:2007ih}.  In particular, we find the predicted cyclotron peak at a frequency proportional to the magnetic field.

Being at strong coupling, however, it is not clear how far the particle or quasiparticle picture stretches, and a quasiparticle interpretation of the hydrodynamical model is questionable. The dual effective near-equilibrium description of the (2+1)-dimensional defect flavor degrees of freedom (dofs) would necessitate the coupling of the hydrodynamics with scalar fields that describe the wobbling of the interface. It is also important to remember that since the defect dofs are treated in the quenched approximation ($N_f\ll N_c$), in the presence of a finite external electric field, the momentum of charged dofs is not conserved and effectively dissipates into the gluonic sector, even in the translationally invariant case. Up to time scales parametrically large in $N_c$ , the ambient plasma, consisting of adjoints of the ${\cal N}=4$ super Yang-Mills sector, continues to approximately sit at rest and acts as a momentum reservoir for charged degrees of freedom  \cite{Karch:2008uy,OBannon:2008xbp}. The longitudinal conductivities can be finite and take nonzero values at vanishing frequency, contrary to \cite{Amoretti:2021fch}. A hydrodynamical description without long time tails therefore seems insufficient. Nevertheless, we attempt to infer lessons of physics interest by matching onto an existing hydrodynamical model at finite magnetic field, namely \cite{Hartnoll:2007ih}. We find that it works surprisingly well, implying that, for example, higher quasi-normal modes are not relevant, at least in the frequency range studied here.

Finally, we investigate the combined effects of a magnetic field and pinning.   As we showed in \cite{Jokela:2017ltu}, a symmetry-breaking magnetic lattice pins the stripes, preventing them from sliding.  The DC conductivity across the stripes then drops by an order of magnitude and a pinning pole at nonzero frequency appears, representing the damped harmonic oscillation of the stripes in the pinning potential.  Surprisingly, the effects of the magnetic field seem to be suppressed by the pinning potential, and the conductivities closely resemble those found in \cite{Jokela:2017ltu} at nonzero magnetic field.  Unlike in \cite{Song:2019rnf} where distinct cyclotron and pinning poles were observed, we find only evidence of a single pole; the cyclotron pole appears to transition into a pinning pole as the lattice is introduced.

The outline of the paper is as follows.  We first review in Sec.~\ref{sec:model} the construction and phases of the model.  In Sec.~\ref{sec:homo}, we describe the conductivity of the homogeneous phase, describing the inverted semi-circle law and temperature dependance of the Hall angle. Then, in Sec.~\ref{sec:stripes}, we compute the conductivity of the inhomogeneous striped phase, both with and without the addition of pinning.  We conclude with a discussion of open problems in Sec.~\ref{sec:Discussion}.

\section{D3-D7' Model}\label{sec:model}

\subsection{Set-up}
\label{sec:setup}
The D3-D7' model is a holographic description of strongly interacting fermions on a $(2+1)$-dimensional defect interacting with a $(3+1)$-dimensional gauge field \cite{Bergman:2010gm}. 
 This construction is a member of the well studied class of \#ND=6  brane intersection models ~\cite{Sakai:2005yt,Sakai:2004cn,Jokela:2011eb,Jokela:2011sw,Jokela:2015aha}, whose low-energy excitations are purely fermionic.  An analysis of the magnetic properties of the D3-D7' model was initiated in \cite{Jokela:2012vn}.\footnote{See also \cite{Bergman:2012na} for a review of magnetic properties of probe-brane models.} This system exhibits the quantum Hall effect, featuring a gapped state at nonzero filling fraction \cite{Bergman:2010gm}.  Furthermore, at large charge density, the ground state is spontaneously striped, with spatially modulated magnetization, persistent transverse currents, and modulated charge density \cite{Jokela:2016xuy}. We will only briefly review the set up of the model here.  For a more thorough treatment, see \cite{Jokela:2016xuy}.

The model consists of a probe D7-brane in a D3-brane background, filling the $t,x$, and $y$ boundary dimensions and the holographic radial direction $r$, and wrapping two 2-cycles in the internal $S^5$.   This embedding completely breaks supersymmetry but can be stabilized by wrapping internal magnetic fluxes on the internal 2-cycles, labeled by parameters $f_1$ and $f_2$.  For definiteness, we will consider $f_1 = f_2 = \frac{1}{\sqrt{2}}$, unless otherwise specified.

The D7-brane action is the sum of a Dirac-Born-Infeld (DBI) term and a Chern-Simons (CS) term:
\be 
\label{totalact}
 S  =  -T_7 \int d^8x\, e^{-\Phi} 
 \sqrt{-\mbox{det}(g_{\mu\nu}+ 2\pi\alpha' F_{\mu\nu})} -\frac{(2\pi\alpha')^2T_7}{2} \int P[C_4]\wedge F \wedge F \ , 
\ee
where $T_7$ is the tension of the D7 brane, $\Phi=0$ for the AdS$_5\times$S$^5$ Schwarzshild background geometry, and $P[C_4]$ is the pullback of potential of the self-dual five-form flux~\cite{Bergman:2010gm}.
Bulk solutions to the equations of motion derived from this action are described by the embedding functions $z(r)$,  the boundary direction in which the defect is localized, and $\psi(r)$, the azimuthal angle on the internal $S^5$, as well as the worldvolume gauge field components $a_x$ and $a_y$.\footnote{We work in radial gauge, so that $a_r = 0$.}  A chemical potential $\mu$ and magnetic field $b$ are included by considering appropriate components of the bulk gauge field $a_\mu$.  We will focus our attention on embeddings with zero fermion mass, and we set the AdS scale to one.

At nonzero temperature $T$, the D3-brane background has a black hole horizon at $r_T = \pi T$.  We  scale out the temperature by rescaling the spatial coordinates $x^\mu$ and gauge field $a_\mu$ by the horizon radius $r_T$.  We also work with a compact radial coordinate 
\be
 u = \frac{r_T}{r} \ ,
\ee
which sets the location of the horizon at $u=1$ and the anti-de Sitter (AdS) boundary at $u=0$.

A nonzero chemical potential $\mu$ induces a boundary charge density $d$.  This charge is distributed among two components with very different physical properties.  There are fractionalized charges located in the bulk at the horizon and cohesive charges in the form of D5-branes smeared in the D7-brane worldvolume. In terms of the physical charge density $D$, 
 the rescaled charge density is 
 $d = 2\sqrt{2}\pi D/(\sqrt{\lambda}N_c T^2)$, 
 and the rescaled magnetic field is $b  =  B/T^2 \cdot 2/(\pi \sqrt{\lambda})$  where $B$ is the physical magnetic field.\footnote{For more general internal fluxes $f_1$ and $f_2$ (which still give the fermions non-anomalous mass dimension), the relation for $d$ in terms of $D$ involves an additional factor of $(\sin(2\psi_\infty))^{-2}$.}


\subsection{Background solutions}
\label{sec:backgrounds}

\subsubsection{Homogeneous phase}

The D3-D7' model has, in general, two classes of homogeneous solutions.  The black hole embeddings, in which the D7-brane crosses the horizon, describe gapless conducting states, and will be the focus of our attention in this paper.  In addition, for certain filling fractions, i.e. ratios of the charge density to the magnetic field, there are Minkowski embeddings, for which the D7-brane smoothly ends outside the horizon, resulting in a gapped insulator. The existence of Minkowski embeddings also relies of the D7-brane fluxes falling within certain ranges.  For the fluxes considered here, $f_1=f_2=1/\sqrt{2}$, only black hole embeddings exist.

At zero temperature and with $b=\mu= 0$, the D7-brane embedding is trivial and the induced metric is $AdS_4 \times S^2 \times S^2$, which is dual to a CFT \cite{Jokela:2012vn}.  This indicates the presence of a quantum critical point (QCP) associated with a second-order phase transition from the homogeneous phase to a spatially modulated striped phase \cite{Bergman:2011rf, Jokela:2014dba}.

The magnetic properties of the homogeneous conducting phase were studied in \cite{Jokela:2012vn}.  The DBI and CS terms in the action \eqref{totalact} generate competing contributions to the magnetization.  At the QCP, the DBI dominates and gives a negative contribution, so the system is diamagnetic.  Moving away from the QCP, for large enough charge, the positive contribution from the CS is more important and the homogeneous phase becomes paramagnetic.  At nonzero fermion mass, the system has a spontaneous magnetization and is therefore a ferromagnet.

\subsubsection{Striped phase}
Above a critical charge density, the ground state is a spontaneously striped phase, with a wavelength $L$ (or equivalently, spatial frequency $k_0 = \frac{2\pi}{L}$).  This spatially modulated state exhibits several types of intertwined spatial order in the form of coupled charge and spin density waves as well as modulated persistent currents.   Without loss of generality we can assume that this striped phase has spatial modulation in only one of the spatial coordinates, $x$, and all fields are independent of the other coordinate, $y$.

An applied magnetic field suppresses the spatial modulation, and for a strong enough field, no striped solution exists.  In \cite{Jokela:2014dba}, a magnetic field was shown to increase the critical charge density and cause the phase transition to become first order.   The magnetic field also spreads the stripes, increasing the wavelength $L$.   

\begin{figure}[!ht]
\center
 \includegraphics[width=0.50\textwidth]{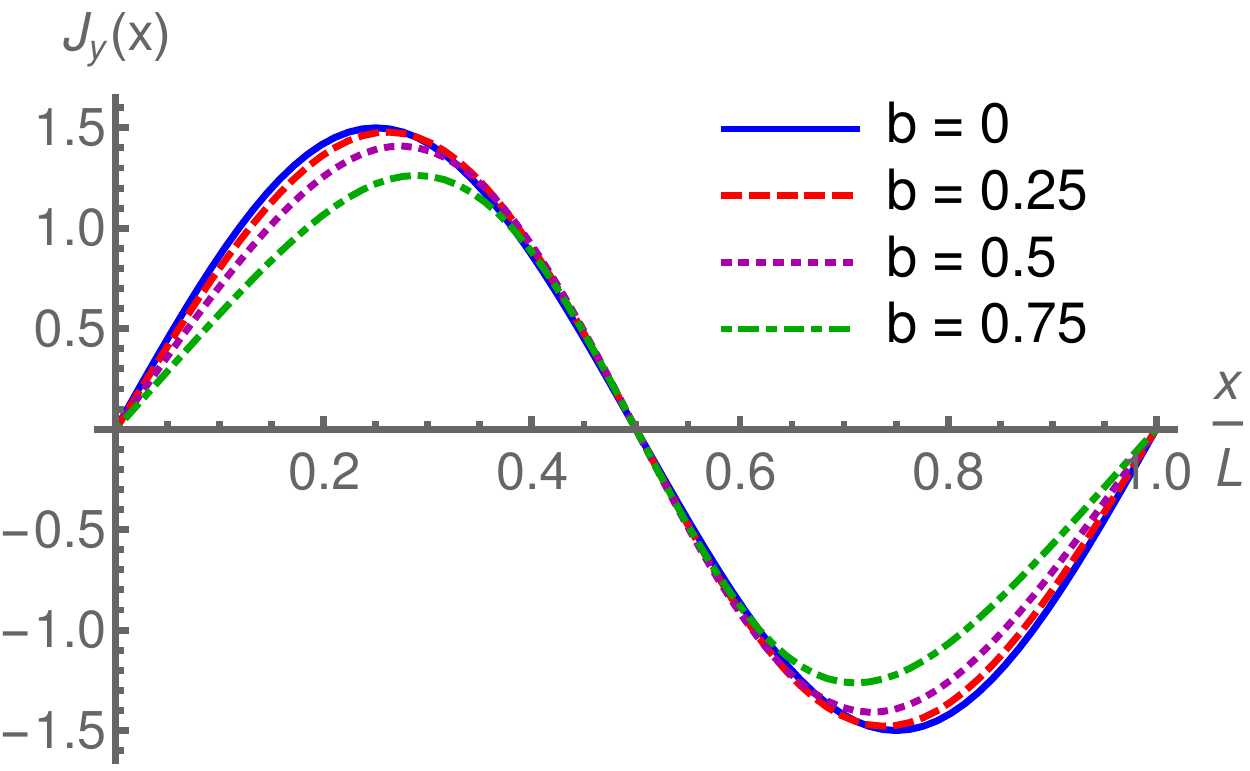}%
  \includegraphics[width=0.50\textwidth]{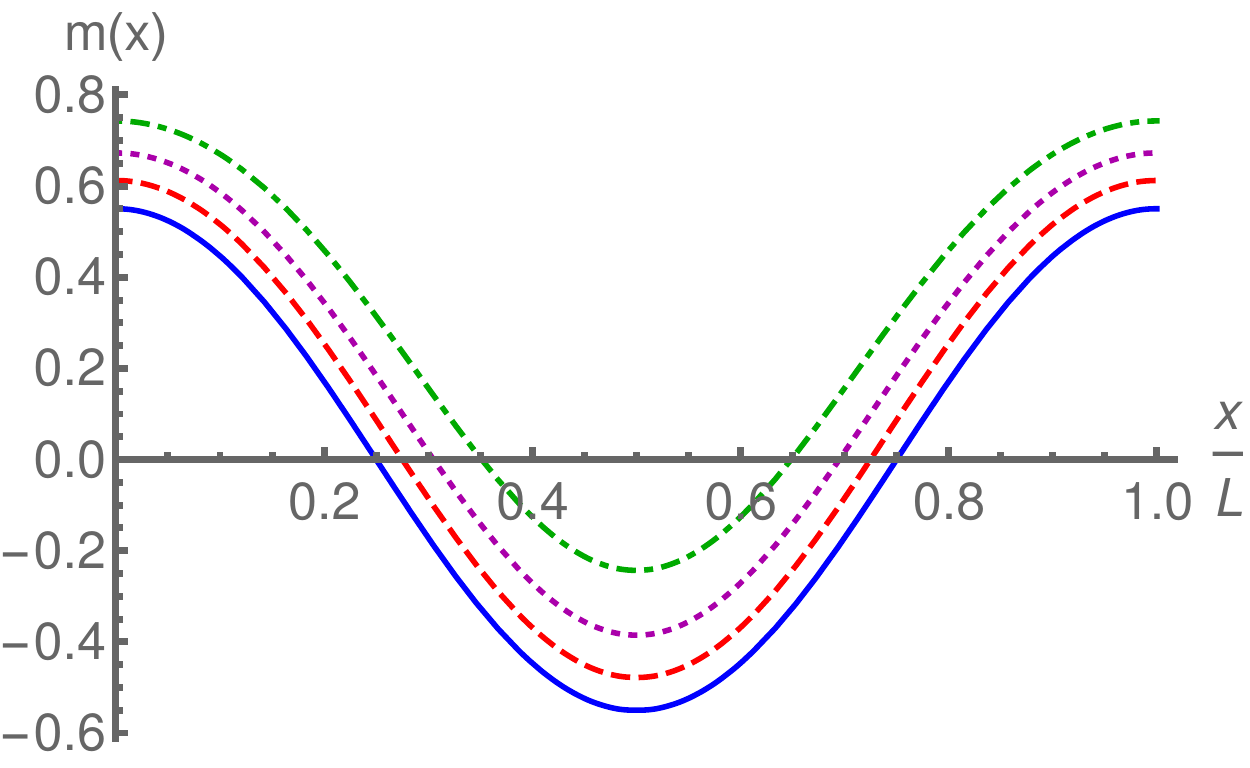}
 \caption{The effect of magnetic field on the striped background at $\mu=4$: (Left) The persistent transverse current  $J_y(x)$, and (Right) the magnetization density $m(x)$ over a spatial wavelength.  The blue, red dashed, magenta dotted, and green dot-dashed curves are the results at  $b=0$, $0.25$, $0.5$, and $0.75$, respectively. We chose units here such that $T_7V_6 L_{AdS}^5 = 1$.}
  \label{fig:bg_magnetic}
\end{figure}

We find the magnetic field reduces the amplitude of the inhomogeneity as well.  For example, the modulated persistent currents flowing along the stripes are suppressed by an applied magnetic field. Fig.~\ref{fig:bg_magnetic} shows the transverse current density $J_y$ as a function of $x$ for increasing $b$.   In addition, the magnetic field breaks the $x \to \frac{L}{2} - x$ parity symmetry, as evidenced by the asymmetric form of the modulation in Fig.~\ref{fig:bg_magnetic} at nonzero $b$.

The magnetization $M$ of the striped phase
\footnote{Notice that the derivatives with respect to $b$ in \eqref{Mdef} and \eqref{magndef} only act on the explicit dependence on the magnetic field, i.e., the field strength components $F_{xy}=-F_{yx}$. Moreover, in the striped phase it is often convenient to define the grand potential as the integral over one period $L$ rather than the full space: $\Omega = - V_6 \int_0^Ldx \int _0^1 du\, \mathcal{L}$. The period $L$ depends on $b$ and is found by minimizing $\Omega/L$~\cite{Jokela:2014dba}. We can now define magnetization as
\be
\hat M = 
\frac{1}{L}\int_0^L dx\, m(x) =  \frac{\partial}{\partial b} \left(\frac{\Omega(T,\mu,b,L)}{L}\right) \ .
\ee
One might be worried that this definition is not consistent with~\eqref{Mdef} and~\eqref{magndef} due to the additional variable $L$ and its dependence on $b$. However since $L$ is the found by minimizing $\Omega/L$ the derivative with respect to $L$ vanishes, so the derivatives of $\Omega/L$ with respect to $b$ at constant $L$ or along $L=L(b)$ are equal.}
 is given by
\be \label{Mdef}
M = -\frac{\partial }{\partial b}\Omega(T,\mu,b) \ ,
\ee
where $\Omega$ is the grand potential energy of the D7-brane, given by the negative of the on-shell action \eqref{totalact}, $\Omega= - S$.  Similar to what was found for the homogeneous phase at large charge \cite{Jokela:2012vn}, the striped phase is paramagnetic.  In addition, the striped phase has a spatially modulated magnetization density $m(x)$, defined by 
\be \label{magndef}
M = \int m(x)\, dx \ , \qquad m(x) = V_6\int_0^1 du\, \frac{\partial \mathcal{L}}{\partial b} 
\ee
where $\mathcal{L}$ is the Lagrangian density of the full probe brane action, $S = \int d^8x \, \mathcal{L}$, and $V_6$ is the volume of $\mathbb{R}^2 \times S^2 \times S^2$, arising from integrations over $t$, $y$, and the angular variables. 
The magnetization density at $\mu=4$ is shown in Fig.~\ref{fig:bg_magnetic}.  As $b$ is increased, it 
increases approximately uniformly, with the modulation remaining approximately constant. In the analysis of the striped phase we will focus on the results at $\mu=4$. This value is large enough for the stripes to be sizeable, but not so large that we would be in the region of asymptotically high densities.

\subsubsection{Pinning}
In the striped phase, spatial translation symmetry is broken spontaneously, resulting in a sliding Goldstone mode.  However, typical experimental systems feature additional sources of explicit translation symmetry breaking, such as a lattice or impurities, which lift the Goldstone mode and effectively pin the stripes.  In the D3-D7' model, two types of pinning potentials were introduced and studied in \cite{Jokela:2017ltu}: a spatially modulated chemical potential models an ionic lattice, and a background antiferromagnetic field represents a magnetic lattice.  

These lattices are implemented holographically via the following boundary conditions for the world volume gauge field:
\bea
{\rm Magnetic \ lattice: } & &a_y(x, u=0) = bx + \alpha_b\sin(k_0 x) \\
{\rm Ionic \ lattice: } & &a_t(x, u=0) = \mu + \alpha_\mu\cos(2k_0 x)
\eea
where $\alpha_b$ and $\alpha_\mu$ give the amplitudes of the two types of lattices. The inhomogeneity of the stripe solutions is driven by the modulation of the bulk gauge field $a_y$~\cite{Jokela:2014dba}.  Consequently, a magnetic lattice couples directly to the stripes, giving rise to a robust pinning effect.  In contrast, the modulation of the temporal component $a_t$ is subleading, causing the pinning effect of the ionic lattice to be suppressed.  In this paper, we will focus on the magnetic lattice because of its direct coupling to the stripes.

In the presence of a pinning potential, the stripes will dynamically adjust so as to be commensurate with the periodicity of the potential, at least if the potential is sufficiently strong.  Here, as in \cite{Jokela:2017ltu}, in order to simplify the computations, we set the wavelength of the lattice to be equal to the dynamically preferred wavelength of the stripes, imposing commensurability by hand.

\section{Conductivity in the homogeneous phase}\label{sec:homo}

\subsection{DC conductivity}


The DC conductivities of the gapless, homogeneous phase were computed in~\cite{Jokela:2012vn}, using the Karch O'Bannon technique \cite{OBannon:2007cex}.  For generic internal fluxes $f_1$ and $f_2$, these can be expressed in terms of horizon data as
\bea
\label{sigmaxx_BH}
\sigma_{xx}^{DC} = \sigma_{yy}^{DC} & = &
\frac{1}{1+b^2} \sqrt{\tilde{d}^2 + \frac{1}{2}
(f_1^2+4\cos^4\psi_T)(f_2^2+4\sin^4\psi_T)(1+b^2)} \\
\label{sigmaxy_BH}
\sigma_{xy}^{DC}  & = & 
\frac{b}{1+b^2}\tilde{d}+
\sqrt{2} 
c(\psi_T) \ ,
\eea
where $\tilde d \equiv d -
\sqrt{2}  
b c(\psi_T)$,
the integration constant $d$ is given by $d = - \partial_u a_t|_{u=0}$,
and
\be
c(\psi) = \psi - \frac{1}{4}\sin\left(4\psi\right) - \psi_0 + \frac{1}{4}\sin(4\psi_0).
\ee
The horizon value of $\psi$ is denoted $\psi_T = \psi(u=1)$, and $\psi_0 = \psi(u=0)$ is the value at the UV boundary.  For the choice of equal fluxes $f_1 = f_2$, the boundary value is $\psi_0 = \frac{\pi}{4}$.

\subsubsection{The inverted semi-circle law}
\label{sec:semi-circle_homogenous}

\begin{figure}[!ht]
\center
 \includegraphics[width=0.70\textwidth]{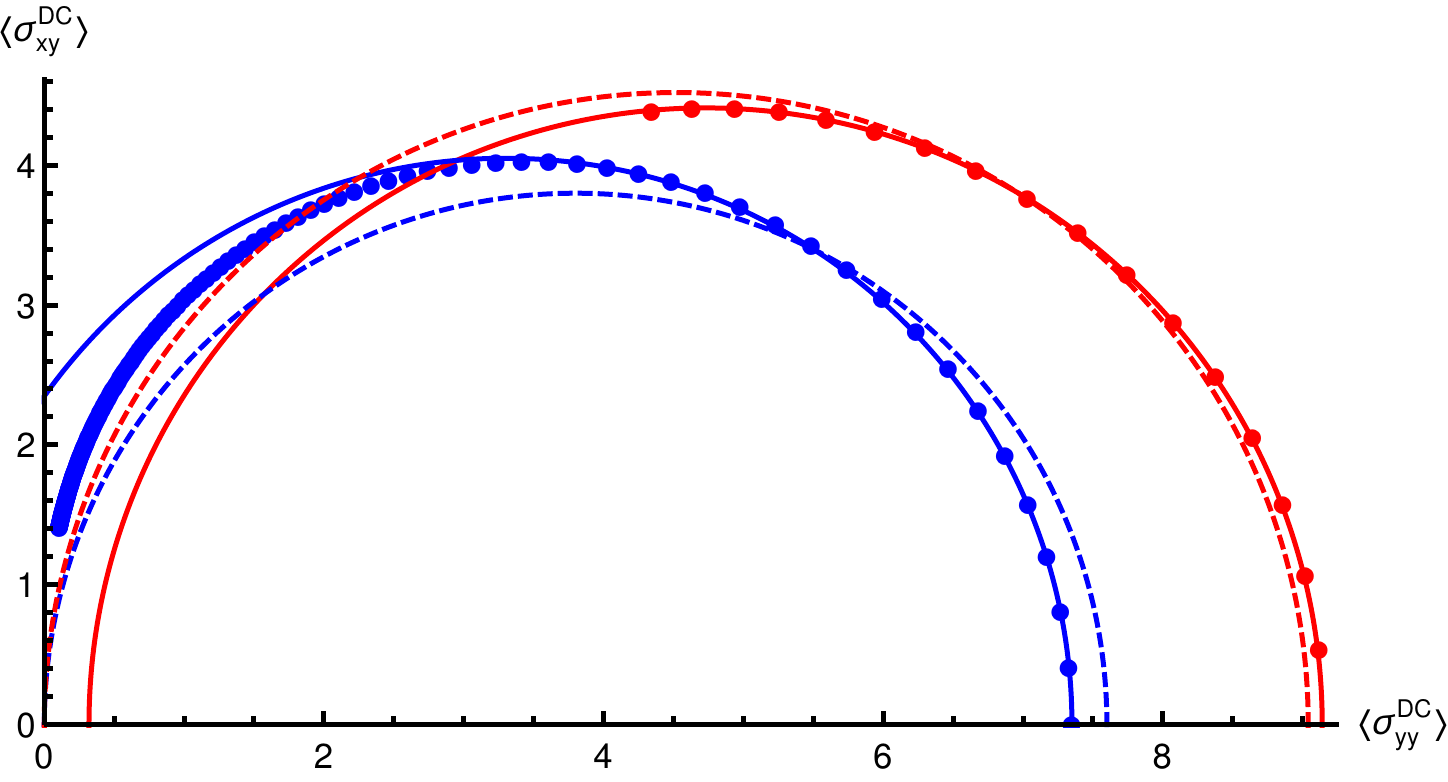}%
 \caption{The semi-circle law for the DC conductivities.  The homogenous phase is shown in blue and the striped phase in red.  For the striped phases the spatially averaged conductivities $\langle \sigma_{xy}^{DC}\rangle$ and $\langle \sigma_{yy}^{DC}\rangle$ are plotted.  The dots are the numerically computed data points at $\mu=4$. The dashed lines illustrate a semi-circular fit constrained to go through the origin, while the solid curve is a semi-circular fit of the low-$b$ data points.} 
  \label{fig:semicircle}
\end{figure}

Plotting the Hall conductivity \eqref{sigmaxy_BH} versus the longitudinal conductivity \eqref{sigmaxx_BH} for different values of $b$,  yields  Fig.~\ref{fig:semicircle}.  We find the result produces, to good approximation, a semi-circle in the $\sigma_{yy}^{DC} - \sigma_{xy}^{DC}$ plane.\footnote{For the homogenous phase,  the choice of which longitudinal conductivity to plot is irrelevant.  However, for the striped phase $\sigma_{xx}^{DC}$ and $\sigma_{yy}^{DC}$ are only approximately equal.  Plotting $\sigma_{xy}^{DC}$ against $\sigma_{xx}^{DC}$ is qualitatively the same, but the semi-circle radius is slightly smaller.  See Sec.~\ref{sec:semi-circle_stripes} for more details.}
 The curve begins at zero magnetic field on the $\sigma_{yy}^{DC}$ axis, and proceeds counterclockwise toward the origin as $b$ is increased.  Interestingly, roles of $\sigma_{xx}^{DC}$ and $\sigma_{xy}^{DC}$ in the semi-circle law are reversed with respect to the law~\eqref{eq:QHsemicircle} which has been found in transitions between quantum Hall states \cite{MacDonald_2000,von_Oppen_2000}.  

The high temperature limit of the DC conductivities was studied in~\cite{Jokela:2012vn}. As it turns out, the limit of low temperatures and small magnetic fields is also interesting. As temperature has been scaled out, this is equivalent to the limit of large charge densities, $d \to \infty$ in our setup. 
At finite charge density, the curve only approximates a semi-circle.  However, in the limit where $d\to \infty$, the semi-circle law becomes exact, as we will now show.

In the large-$d$ limit, the equation of motion for the embedding scalar $\psi(u)$ simplifies to 
\be
 \partial_u\left(\frac{h(u)\psi'(u)}{1+u^2h(u)(\psi'(u))^2}\right) = \morder{\frac{1}{d}} \ ,
\ee
where $h(u)=1-u^4$ is the blackening factor.
Therefore the regular solution is simply constant up to subleading corrections:
\be \label{eq:psilarged}
\psi(u) = \psi_\infty + \morder{\frac{1}{d}} \ ,
\ee
with the value of the constant determined by the UV behavior.\footnote{This conclusion may appear premature because the solution may in principle have irregular behavior near the boundary or near the horizon which could alter the value of the constant, which is not captured by the simple minded analysis where $u$ does not scale with $d$. A careful analysis of the subleading corrections however verifies that~\eqref{eq:psilarged} is correct.} 
Substituting $\psi_T = \psi_\infty + \morder{1/d}$ in to equations~\eqref{sigmaxx_BH} and \eqref{sigmaxy_BH}, and using the fact that $c(\psi_\infty) =0$, we find\footnote{Notice that the leading-order result holds regardless of  the horizon value $\psi_T$, but the condition $\psi_T = \psi_\infty + \morder{1/d}$ ensures that the \order{d^0} corrections vanish. Moreover inserting the constant solution for $\psi$ in the gauge field equation of motion, and integrating, we find 
\be
 \mu = \frac{ \Gamma \left(\frac{1}{4}\right) \Gamma \left(\frac{5}{4}\right) \sqrt{d}}{
 \sqrt{\pi }\, \sin( 2 \psi_\infty)} +\morder{\frac{1}{\sqrt{d}}}\ .
\ee}
\bea
\sigma_{xx}^{DC} & = & 
\frac{d}{1+b^2} + \morder{\frac{1}{d}}  \label{sigmaxx_BH_ld} \\
\sigma_{xy}^{DC} & = & 
\frac{b d}{1+b^2} + \morder{\frac{1}{d}} \label{sigmaxy_BH_ld} \ .
\eea

In this high-density limit, the leading-order conductivities, \eqref{sigmaxx_BH_ld} and  \eqref{sigmaxy_BH_ld},  satisfy an exact semi-circle law:
 \be
 \label{eq:semicircle_homogenous}
  \left(\sigma_{xx}^{DC}-\frac{d}{2}\right)^2 + \left(\sigma_{xy}^{DC} \right)^2 = \left(\frac{d}{2}\right)^2 \ . 
 \ee
The conductivities trace out a complete semi-circle of radius $d/2$ as $b$ varies from zero to infinity, with the maximal value of $\sigma_{xy}^{DC}$ obtained at $b=1$.  The resistivity matrix takes a very simple form:
 \be
  \left(\begin{array}{cc}
   \rho_{xx} & \rho_{xy} \\
   \rho_{yx} & \rho_{yy} 
  \end{array}\right) 
   = \frac{1  }{d}\left(\begin{array}{cc}
   1 & -b \\
   b  & 1 
  \end{array}\right) \ .
 \ee

As these results hold for generic values of the fluxes $f_i$, they may seem to  be in tension with the quantum Hall states which have been found in the model~\cite{Bergman:2010gm}. However, here we are  taking $d \gg b$, which takes us far away from the possible QH states which have fixed filling fractions $\propto d/b$. At fixed large $d$, the regime of QH states is entered for $b \sim d$, i.e., near the origin in the conductivity plane, where the above approximations fail. 
 
It is worth noting that the semi-circle observed here has some important differences with the usual law seen in quantum Hall transitions.  The usual semi-circle law has a fixed density-independent radius and is observed at low temperature and with charge density and magnetic field of the same order, at a filling fraction of order one.  Here,  the radius scales as the physical charge density over the temperature squared, and the limit of large $d$ but finite $b$ corresponds to large filling fraction.

Note also that  the limit $d \gg b$ is well within the regime where striped instabilities are present (for appropriate choices of the fluxes $f_i$). We will study in Sec.~\ref{sec:semi-circle_stripes} below how this semi-circle law is modified in the striped phase.

\subsubsection{Hall angle}

\begin{figure}[!ht]
\center
 \includegraphics[width=0.70\textwidth]{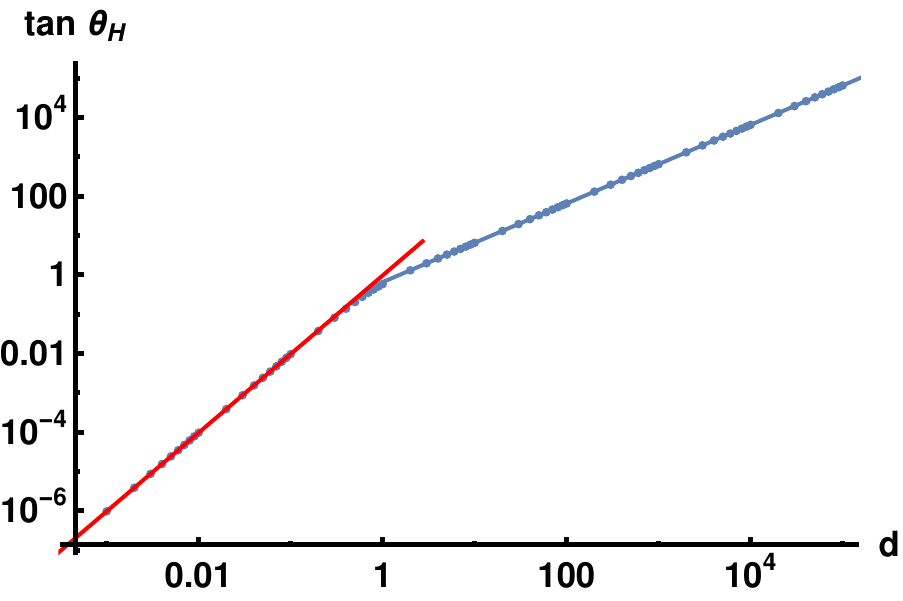}%
 \caption{A log-log plot of the tangent of the Hall angle as a function of $d$, with fixed $d/b = 2$. The dots are numerically computed data points.  The blue line shows the scaling at large $d$, corresponding to low temperatures, $\tan\theta_H \sim d \sim T^{-2}$.  At small $d$, which is high temperature, $\tan\theta_H \sim d^2 \sim T^{-4}$, as shown by the red line.} 
  \label{fig:Hall_angle}
\end{figure}

Using equations \eqref{sigmaxx_BH} and \eqref{sigmaxy_BH}, the Hall angle $\theta_H$, defined by 
\be
\label{Halldefinition}
\tan \theta_H = \frac{\sigma_{xy}^{DC}}{\sigma_{xx}^{DC}} \ ,
\ee
 can be computed numerically.  Both the rescaled charge $d$ and magnetic field $b$ scale with temperature as $T^{-2}$.  Because the temperature has been scaled out,  the temperature can be changed by varying $d$ and $b$ while holding the physical charge density and magnetic field fixed, which operationally means keeping $d/b$ fixed.  The result, shown in Fig.~\ref{fig:Hall_angle}, shows different scaling regimes at high and low temperatures.

The high-temperature limit, which was addressed in \cite{Jokela:2012vn}, corresponds to $d$ and $b$ both going to zero with $d/b$ held fixed.  In this case, $\psi_T \to \psi_\infty$ and $c(\psi_T) \sim d^2$, and so the Hall conductivity goes to zero as $\sigma_{xy}^{DC} \sim d^2 \sim T^{-4}$.  However, the longitudinal conductivity goes to a temperature-independent constant:
\be
\sigma_{xx}^{DC} =  \frac{1}{\sqrt{2}} \sqrt{(f_1^2+4\cos^4\psi_\infty)(f_2^2+4\sin^4\psi_\infty)}  \ .
\ee
Together this gives
\bea
\tan \theta_H \sim T^{-4} \ .
\eea

The low-temperature limit, which is of more interest for comparison with experiments, corresponds to the opposite: $d$ and $b$ both going to infinity with $d/b$ held fixed.  In this limit, $c(\psi_T)$ is constant, and so the Hall conductivity \eqref{sigmaxy_BH} is
\bea
\sigma_{xy}^{DC} & = & 
\frac{\tilde d}{b} + 
\sqrt{2} 
c(\psi_T) 
\\
& = & 
\frac{d}{b}  \ ,
\eea
which is temperature independent.  The longitudinal conductivity \eqref{sigmaxx_BH} now scales as $\sigma_{xx}^{DC} \sim 1/b \sim T^2$, which gives
\bea
\tan \theta_H \sim T^{-2} \ .
\eea
The $T^{-2}$ scaling at low temperature matches the measured scaling in strange metals.\footnote{A $T^{-2}$ scaling was obtained in the Hall angle in a bottom-up holographic model featuring a similar DBI + CS action to the one used here, but with a specific choice of Lifshitz scaling and anisotropic scaling exponent \cite{Pal:2010sx}.}  Note, however, that we do not obtain the signature $T^{-1}$ scaling for $\sigma_{xx}^{DC}$.   The underlying physics can be best understood by looking at the individual longitudinal and Hall conductivities.  Taking their ratio may serve to obscure rather than clarify the physics.  

\subsection{Optical Conductivity}
\label{sec:optical_conductivity_homogeneous}

The optical conductivity can be computed by analyzing the fluctuations on top of the background solutions:  The D7-brane gauge fields are perturbed, and the equations of motion for the linear fluctuations are solved numerically with appropriate boundary conditions.  From these solutions, the conductivity can be extracted.   Further details are given in \cite{Jokela:2016xuy}. The homogeneous phase is also isotropic, so without loss of generality, we consider an applied electric field in the $x$ direction and compute the resulting longitudinal current $j_x$ and Hall current $j_y$.

The optical conductivity of the homogeneous phase at zero magnetic field is in many ways similar to what was found for the striped phase in \cite{Jokela:2016xuy}.  The longitudinal conductivity $\sigma_{xx}$ exhibits a Drude peak around zero frequency.  The width of this peak is due to the dissipation provided by the background D3-branes, which act like smeared impurities.  The Hall conductivity $\sigma_{xy}$ vanishes at $b=0$, as required by parity conservation.

There are, however, some important differences between the conductivities of the striped and homogeneous phases.
In the striped phased, the Hall current along the stripes $\sigma_{yx}$ had a nonzero spatially modulation but the spatial average vanished.  It also featured an additional spatially-modulated delta function at zero frequency.  This was due to persistent transverse currents in the striped background which slide along with the stripes as a result of an applied electric field.  The homogeneous phase lacks these persistence currents, and, not surprisingly, the delta function is absent in the Hall conductivity, even at nonzero $b$.

\begin{figure}[!ht]
\center
 \includegraphics[width=0.50\textwidth]{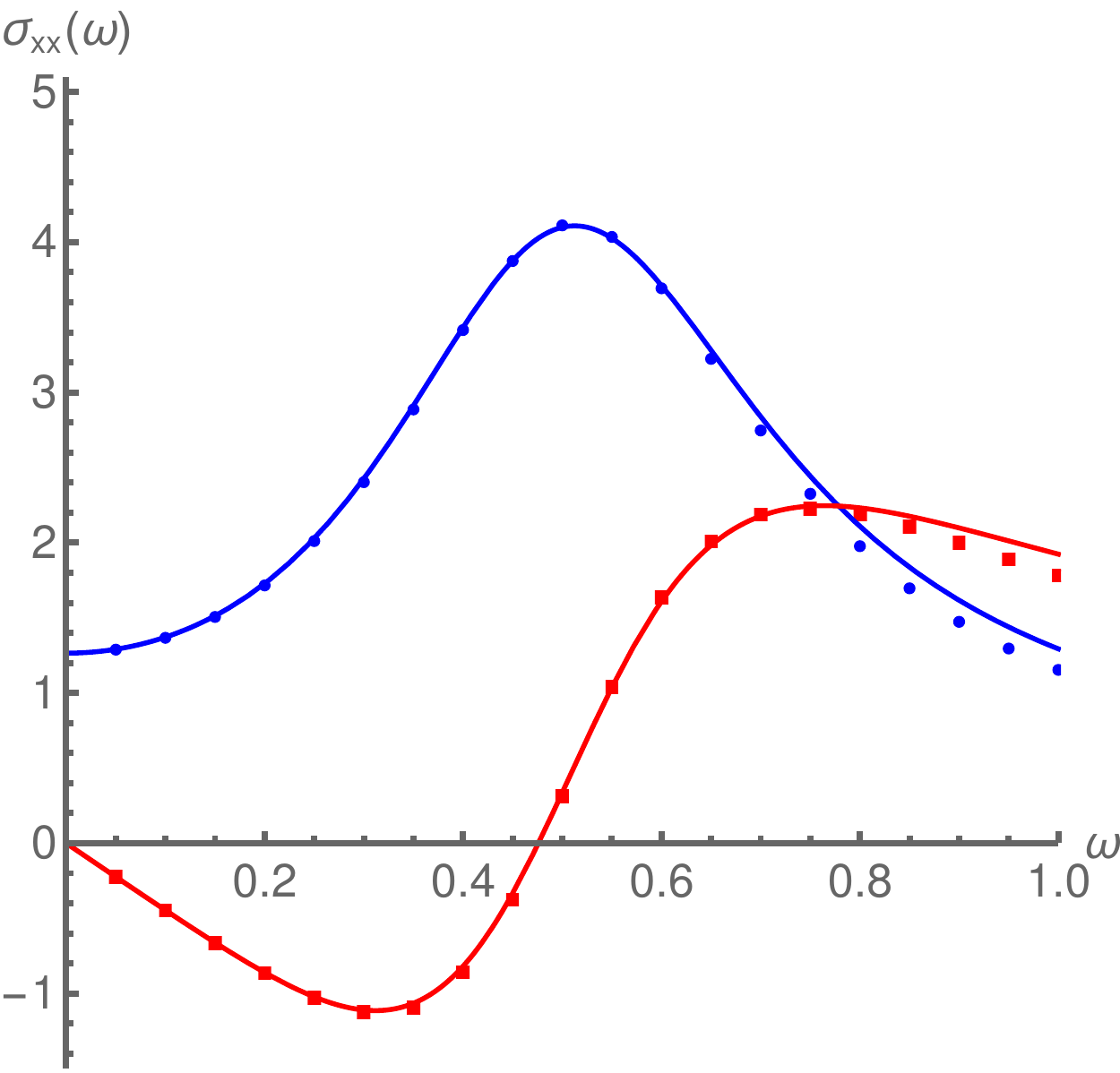}%
 \includegraphics[width=0.50\textwidth]{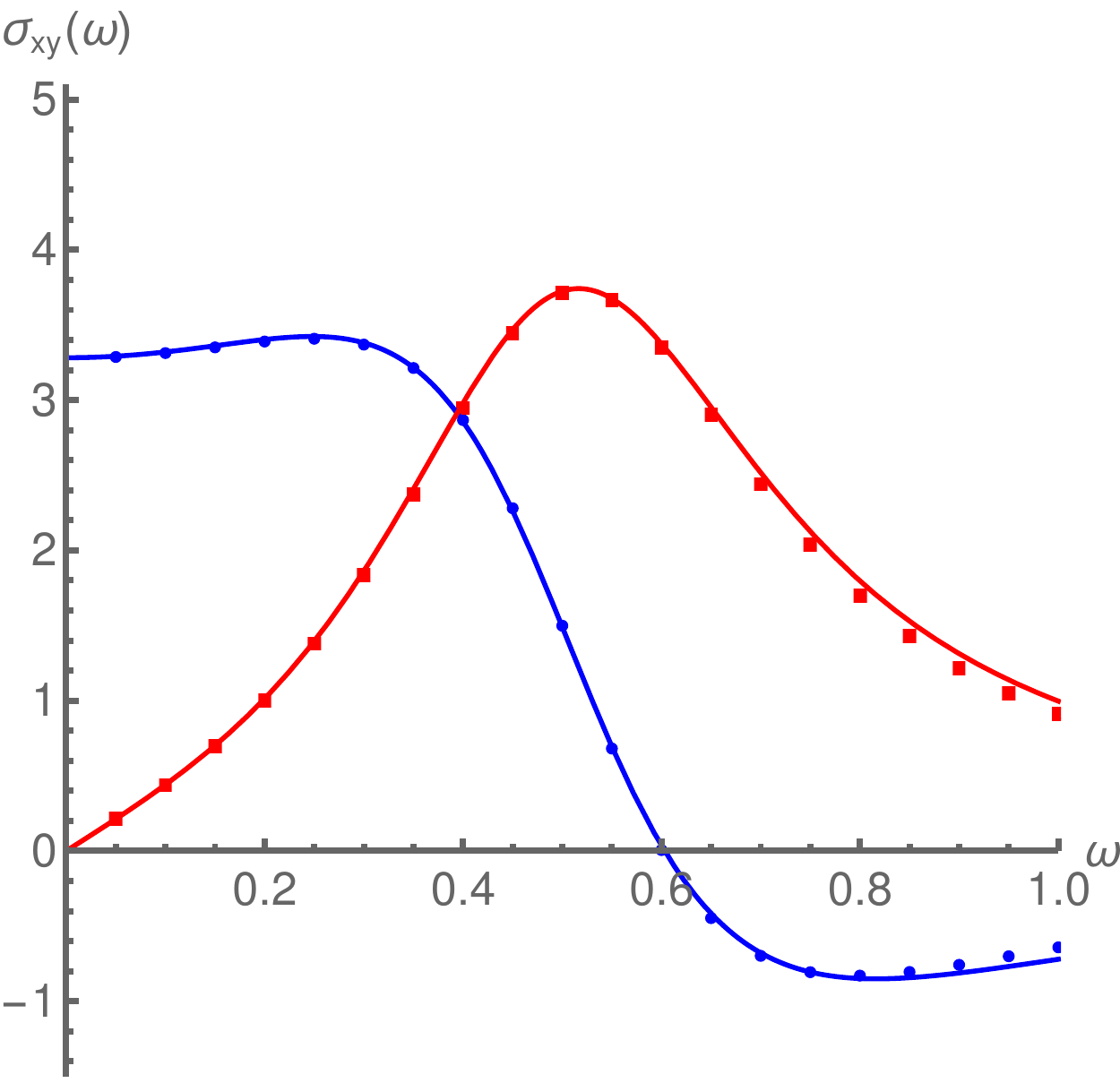}
 \caption{The (left) longitudinal and (right) Hall optical conductivities of the homogeneous phase ($\mu = 4$) with $b=2$, with the real part shown in blue and the imaginary part in red. Dots are the numerically computed points, and the curves show the fits to \eqref{Hartnoll_sigmaxx} and \eqref{Hartnoll_sigmaxy}.} 
  \label{fig:unpinned_fit_homog} 
\end{figure}

When a magnetic field is turned on, the peak in the conductivity moves to higher frequency.  The location of this peak $\omega_c$ is found to be nearly proportional to $b$ and can be interpreted as the cyclotron frequency of the charge carriers.  As shown in Fig.~\ref{fig:unpinned_fit_homog} (left), our results match to good accuracy the hydrodynamical model of \cite{Hartnoll:2007ih}.  At low frequency, the longitudinal conductivity can be fit by\footnote{See Eq.~(3.37) in \cite{Hartnoll:2007ih}.}
\be
\label{Hartnoll_sigmaxx}
 \sigma_{xx}(\omega) = \sigma_Q \left[\frac{\left(\omega+i/\tau\right)\left(\omega+i\gamma+i\omega_c^2/\gamma+i/\tau\right)}{\left(\omega+i\gamma+i/\tau\right)^2-\omega_c^2}\right] \ .
\ee
Here, $\tau$ is the momentum dissipation time due to scattering off impurities.  The quantity $\gamma$ is interpreted as the cyclotron damping due to collisions of current carriers and their antiparticles.  This cyclotron damping frequency is predicted to scale with magnetic field as $\gamma \propto b^2$.  To make the $b$-dependence explicit, we write $\omega_c = \kappa_\omega b$ and $\gamma = \hat \gamma b^2$. Rather than numerically fitting our results to $\hat\gamma$, we use the damping frequency due to vortices $\gamma_V = \omega_c^2/\gamma = \kappa_\omega^2/\hat\gamma$  because it is less sensitive to the details of the fit.   We fix the parameter $\sigma_Q$ by requiring that the DC conductivity $\sigma_{xx}(0)$ is given by the results obtained from~\eqref{finalDCsigmaxx} after inserting the background geometry and the sliding speed $\hat v_{x}$.

In the limit of zero magnetic field, \eqref{Hartnoll_sigmaxx} reduces to the usual Drude form plus a constant
\be
\label{Drude}
 \sigma_{xx}(\omega) = \sigma_Q+ \frac{\sigma_D}{1- i\omega \tau} \ ,
\ee
where the height of the Drude peak is given by $\sigma_D =  \sigma_Q \kappa_\omega^2 \tau/\hat \gamma =  \sigma_Q \tau  \gamma_V$.

Our results for the optical Hall conductivity are shown in Fig.~\ref{fig:unpinned_fit_homog} (right).  These also closely match the hydrodynamical model of~\cite{Hartnoll:2007ih}
\be
\label{Hartnoll_sigmaxy}
 \sigma_{xy}(\omega)= -\sigma_{yx}(\omega) =  - \frac{\rho}{b} \left[\frac{\left(\omega_c^2+\gamma^2-2i\gamma(\omega+i/\tau)\right)}{\left(\omega+i\gamma+i/\tau\right)^2-\omega_c^2}\right] \ ,
\ee
where the parameters $\tau$, $\omega_c$, and $\gamma$ are the same as for the longitudinal conductivity.  However, for the Hall conductivity, the overall coefficient $\rho$, rather than $\sigma_Q$, is determined from the result for the DC conductivity.  The small-$b$ limit also gives a Drude-like form 
\be
\label{Drude_Hall}
 \sigma_{xy}(\omega) =  \frac{2\hat\gamma\rho\tau b}{1- i\omega \tau} - \frac{\rho \kappa_\omega^2 b}{(\omega + i/\tau)^2}  \ ,
\ee
which vanishes in the $b=0$ limit as required by parity conservation. 

\begin{figure}[!ht]
\center
 \includegraphics[width=0.50\textwidth]{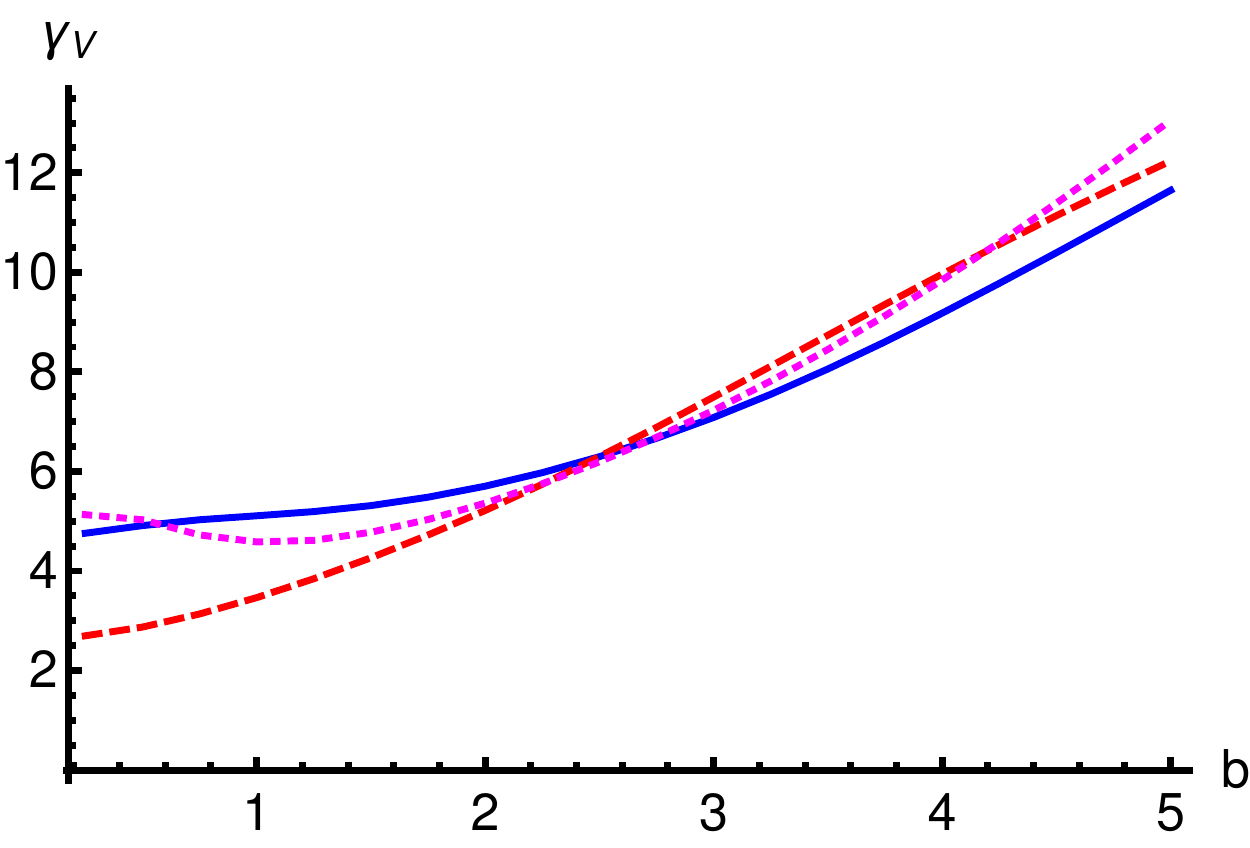}%
 \includegraphics[width=0.50\textwidth]{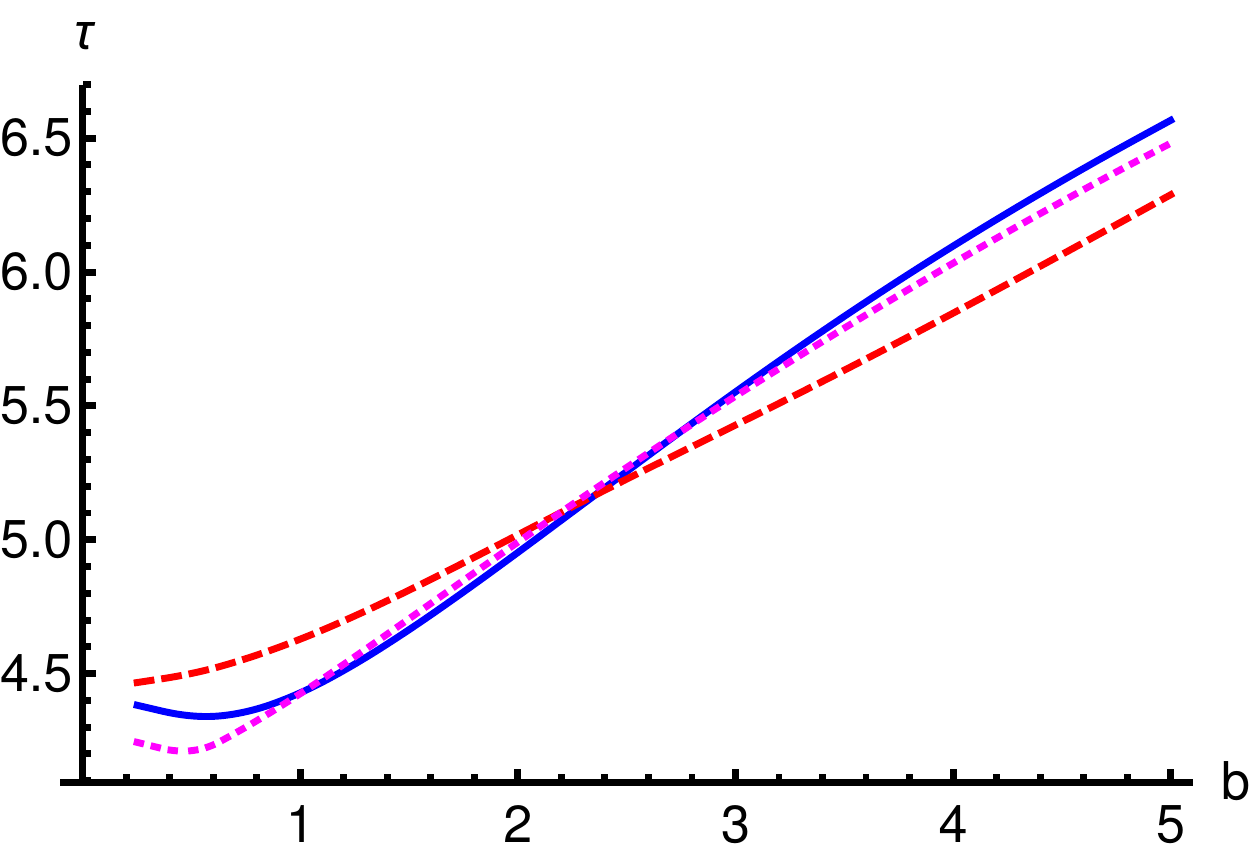}
 \includegraphics[width=0.50\textwidth]{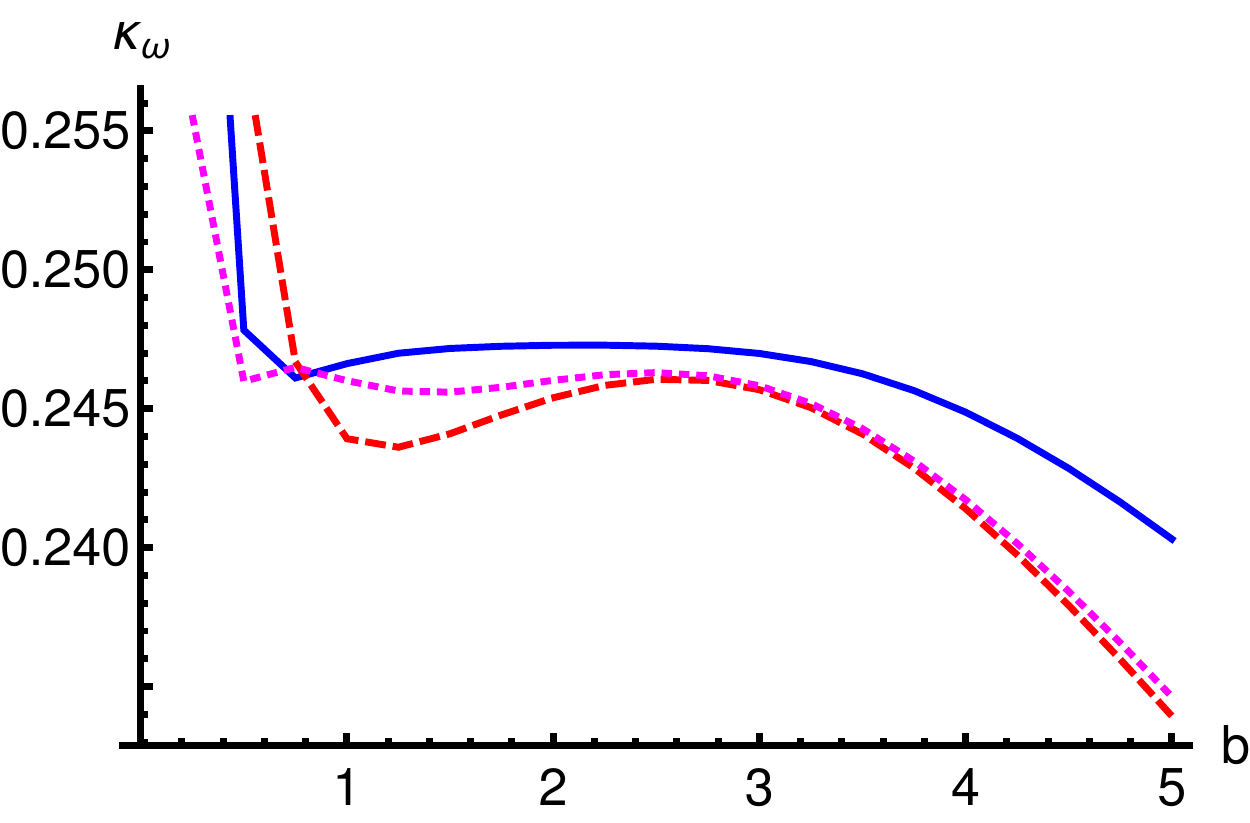}%
 \includegraphics[width=0.50\textwidth]{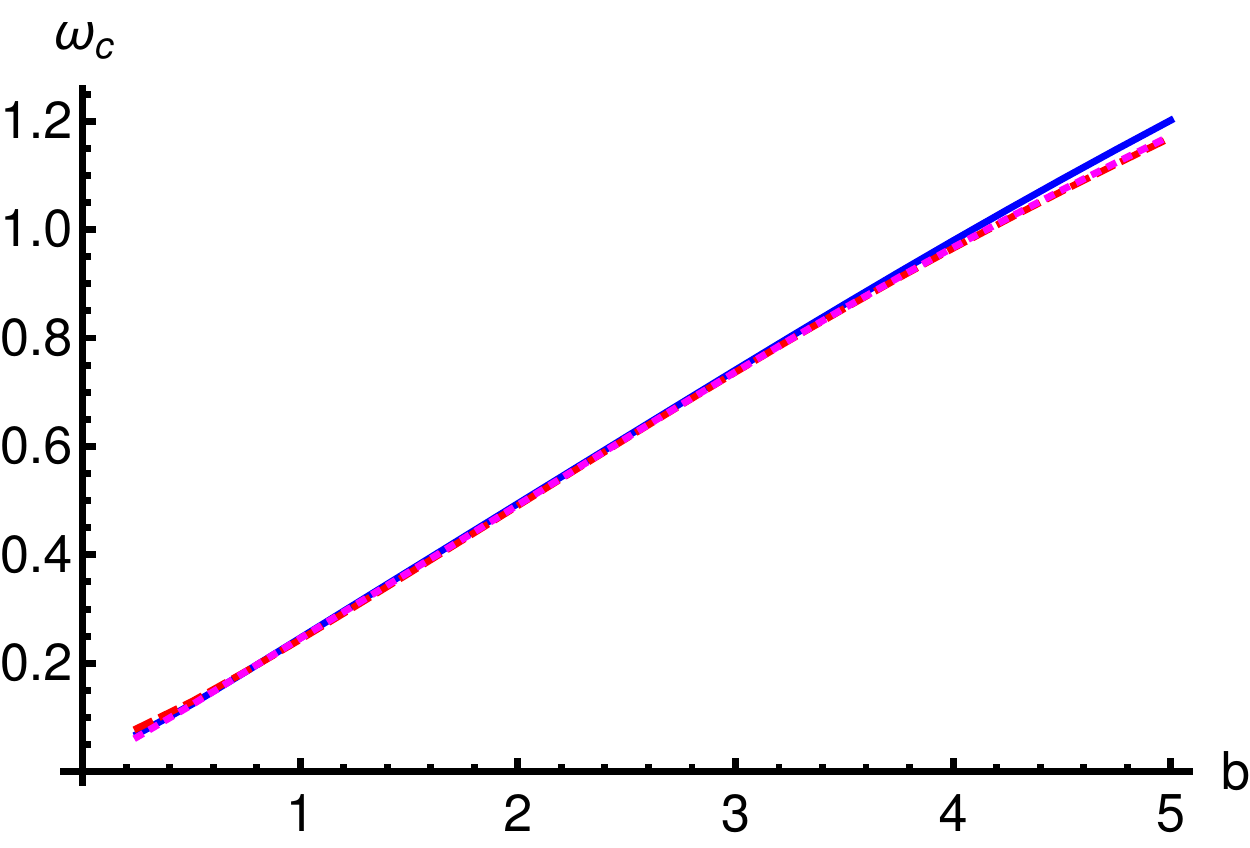}
 \caption{The three fit parameters ($\gamma_V$, $\tau$, $\kappa_\omega$) and the cyclotron frequency $\omega_c = \kappa_\omega b$ for homogeneous optical conductivities as a function of $b$. The solid blue, dashed red, and dotted magenta curves were fitted to the data for the longitudinal conductivity, the Hall conductivity, and to both the conductivities simultaneously, respectively.}  
  \label{fig:fit_params_homog}
\end{figure}

After fixing $\sigma_Q$ and $\rho$, there are three parameters $\gamma_V$, $\tau$, and $\kappa_\omega$, which we numerically fit to \eqref{Hartnoll_sigmaxx} and \eqref{Hartnoll_sigmaxy}.  The curves in Fig.~\ref{fig:unpinned_fit_homog} show the fit for $b=2$, 
which uses data points with $\omega \le 0.65$.
In Fig.~\ref{fig:fit_params_homog}, we show the dependence of the fit parameters on $b$. After scaling out the expected scaling of $\omega_c$ and $\gamma$ with $b$, the fit parameters still exhibit a mild $b$ dependence. The sharp growth of $\kappa_\omega$ at small $b$ is, however, a numerical artifact.  The conductivity becomes insensitive to $\kappa_\omega$ at small $b$ so $\kappa_\omega$ is poorly determined by the fit there.

The fit using the hydrodynamic model in Fig.~\ref{fig:unpinned_fit_homog} is arguably good, and the dependence on the magnetic field of the various parameters in Fig.~\ref{fig:fit_params_homog} appears natural. But since our fits have a large number of parameters, despite their high quality, it is not obvious that they confirm the validity of the hydrodynamic model. In order to analyze this in more detail, we have considered more general fits with a leading pair of quasi normal modes, i.e., fits having two poles in the complex plane
\be \label{eq:polefit}
 \sigma = c + \frac{A_r+iA_i}{\omega-\omega_c+i\Gamma}-\frac{A_r-iA_i}{\omega+\omega_c+i\Gamma}
\ee
both for longitudinal and for Hall conductivities. We do not present the results here since they do not differ significantly from those given by the hydrodynamic model: the fit quality is improved due to the additional parameters, but key parameters such as the cyclotron frequency are essentially unchanged. We find that the additional terms introduced in~\eqref{eq:polefit} with respect to the hydrodynamic model are essentially fit to zero, but in order to see this clearly, one needs to go to relatively high values of the magnetic field (around $b=2$ or $b=3$).\footnote{At smaller values of $b$, the two poles are very close to each other due to the small cyclotron frequency, and the detailed structure near poles cannot be resolved.} Therefore, the results from the more general fit  support the validity of the hydrodynamic model, even at high values of the magnetic field.

\section{Inhomogeneous conductivity}\label{sec:stripes}

\subsection{DC conductivity: sliding stripes}
\label{sec:DCunpinned}

An electric field applied to the striped phase causes the stripes to move.  This sliding mode is precisely the Goldstone mode for the spontaneously broken translation symmetry.  In \cite{Jokela:2016xuy}, the electrical conductivities were derived at vanishing magnetic field.  The stripes were found to slide at a constant velocity proportional to the electric field and carry a significant fraction of the current.

Here, we generalize the computation of the DC conductivity to nonzero magnetic field.  As in \cite{Jokela:2016xuy}, a conserved bulk quantity allows the current across the stripes (i.e. in the $x$ direction) to be computed in terms of analytic expressions of the horizon values of the striped solution.  For the current parallel to the stripes (i.e. in the $y$ direction), there is no analogous conserved bulk quantity, and only spatially-averaged DC conductivities can be computed in terms of horizon data.  The details of the computation are relegated to Appendix \ref{app:DCderivation}, with the expressions for the components of the DC conductivity given in \eqref{finalDCsigmaxx}, \eqref{finalDCsigmaxy}, \eqref{finalDCsigmayy}, and \eqref{finalDCsigmayx}.

At nonzero $b$, an electric field applied in the $x$ direction still causes the stripes to slide.  However, in the presence of a magnetic field, the stripes also slide due to an electric field applied in the $y$ direction, which we term ``Hall sliding."  As defined in \eqref{slidingspeedapp}, the total sliding speed, to linear order in the electric fields $E_x$ and $E_y$ is
\be
\label{slidingspeed}
 v_s = \hat v_x E_x + \hat v_y E_y \ ,
\ee
where the coefficients $\hat v_x$ and $\hat v_y$ are independent of the electric field.  The sliding speed is not determined as part of the computation described in Appendix \ref{app:DCderivation} but must be obtained numerically from the DC limit of the fluctuations.
As was the case in \cite{Jokela:2016xuy}, we extracted the sliding speed by comparing the $x$-derivatives of the background fields $\psi$ and $a_y$, for which the modulation is strongest, to the fluctuation solutions of the same fields, extrapolated to $\omega = 0$.  Alternatively, because modulated persistent currents slide along with the stripes, the sliding speed can also be obtained from the coefficients of the zero-frequency delta functions of the optical conductivities $\sigma_{yx}$ and $\sigma_{yy}$ (see Sec.~\ref{sec:stripesAC} for more details).

\begin{figure}[!ht]
\center
 \includegraphics[width=0.70\textwidth]{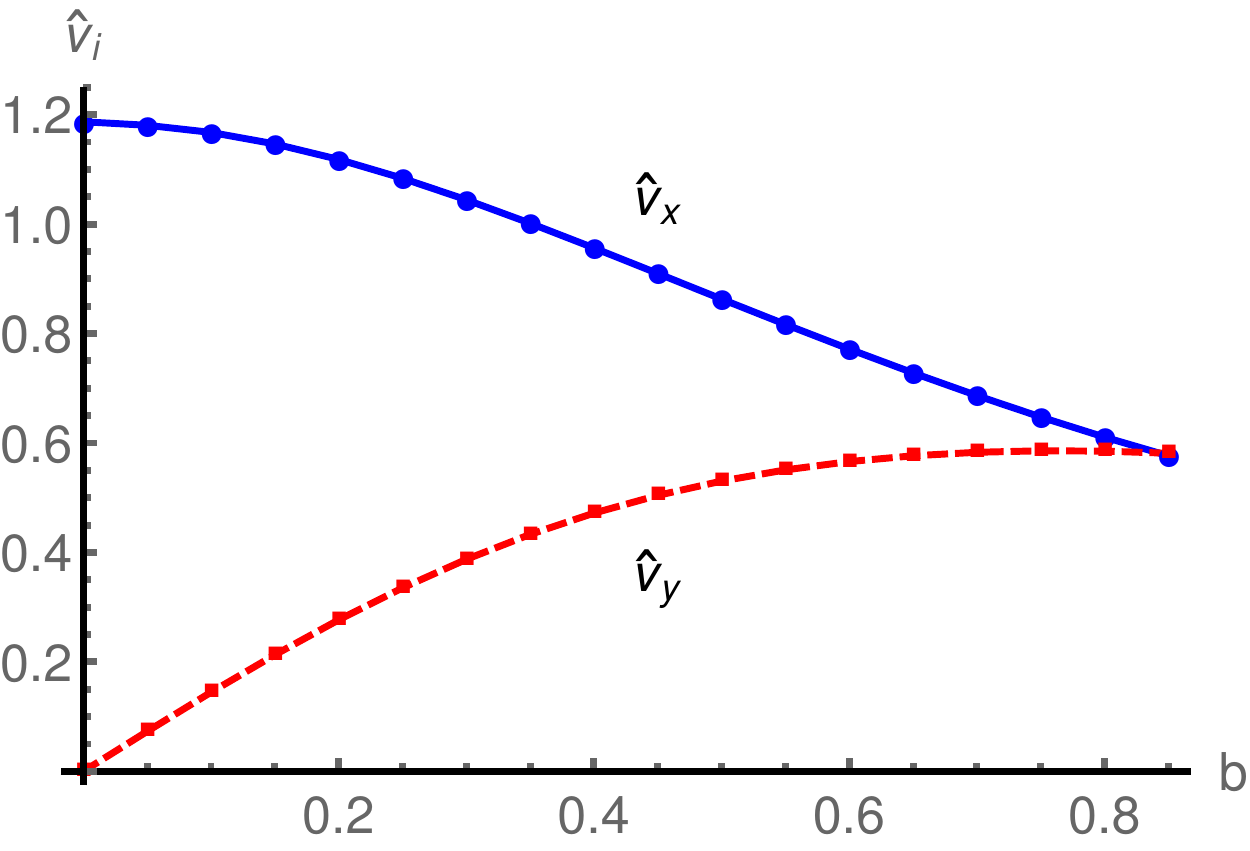}%
 \caption{The sliding speed coefficients $\hat v_x$ (solid blue curve) and $\hat v_y$ (dashed red curve) as functions of magnetic field $b$ for $\mu = 4$.  The  curves were extracted directly from the fluctuations of $\psi$ and $a_y$. As a cross-check, the dots were extracted independently from the weights of the delta functions of the optical conductivities.  For $b > 0.864$, the striped phase is no longer thermodynamically stable, and for $b > 0.95$, no striped solution exists.} 
  \label{fig:vs}
\end{figure}

Our results for $\hat v_x$ and $\hat v_y$ are shown in Fig.~\ref{fig:vs}.  The magnetic field $b$ ranges from zero to $b=0.864$, at which point there is a first-order phase transition to the homogeneous phase.  A thermodynamically unstable striped phase only exists up to $b=0.95$. The two methods of obtaining the sliding speeds, indicated by the curves and the dots, closely agree.  For small $b$, the Hall sliding speed $\hat v_y$ increases linearly before leveling off, while $\hat v_x$ decreases substantially with $b$. 

\begin{figure}[!ht]
\center
 \includegraphics[width=0.50\textwidth]{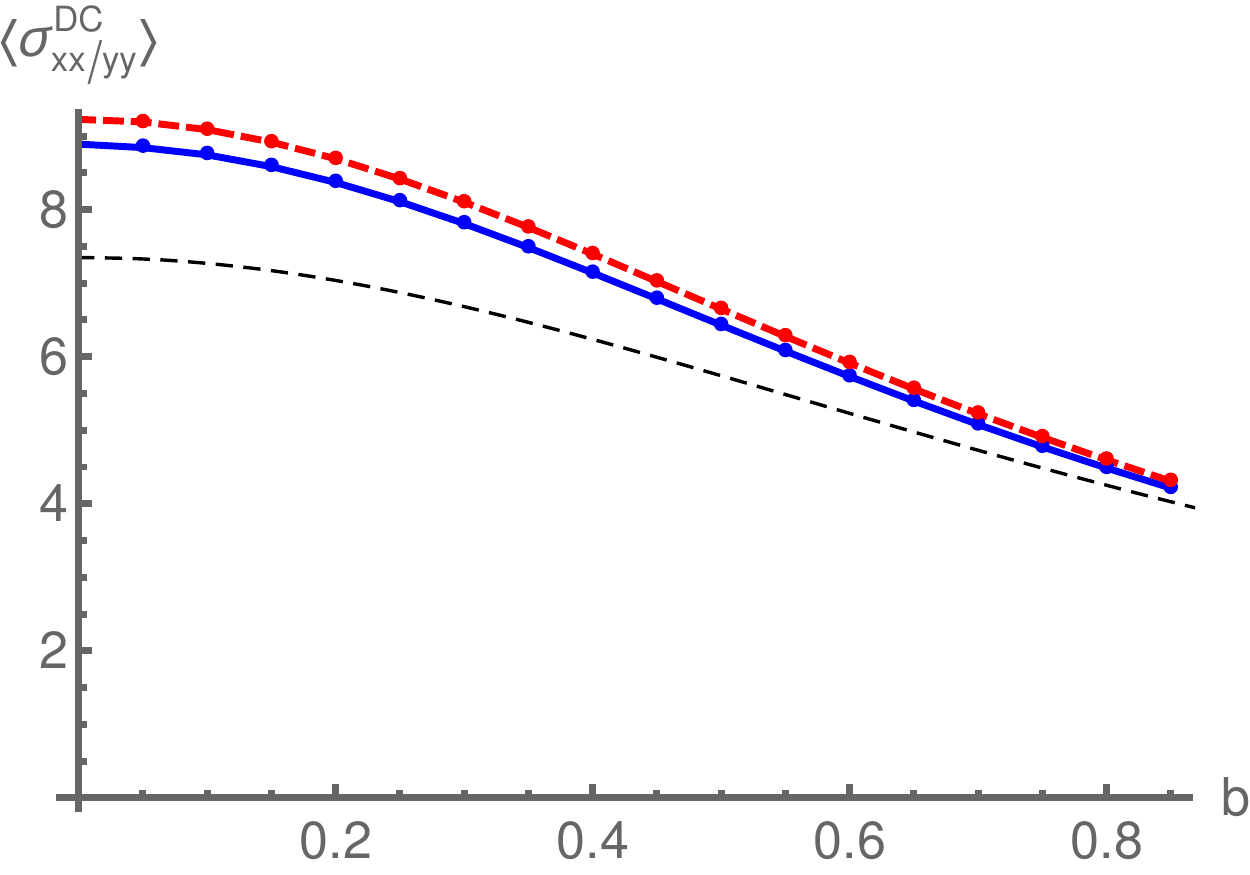}%
 \includegraphics[width=0.50\textwidth]{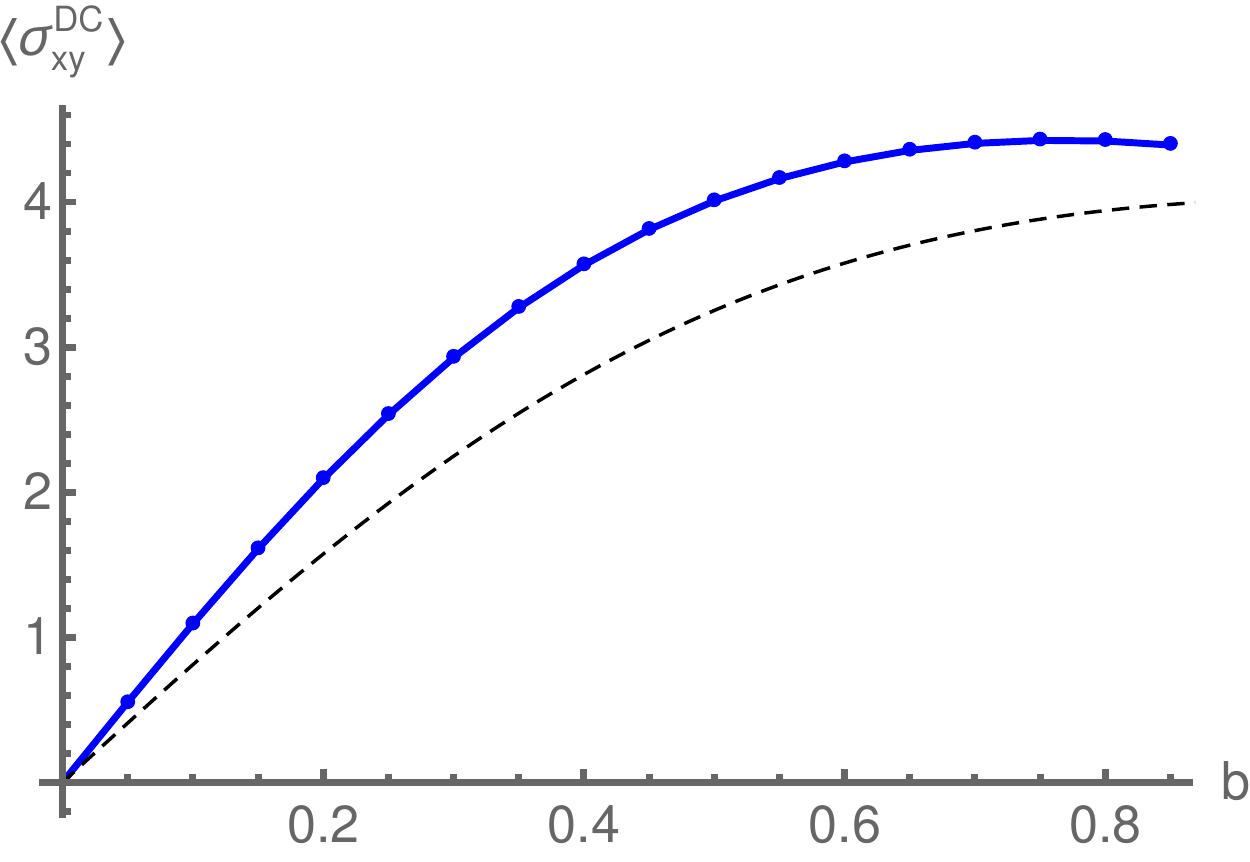}
 \caption{The averaged DC conductivities as functions of the magnetic field for $\mu = 4$: (Left) $\langle \sigma_{xx} \rangle$ is dashed red and $\langle \sigma_{yy}\rangle$ is solid blue; (Right) $\langle \sigma_{xy}\rangle$ and $-\langle \sigma_{yx}\rangle$, whose values are equal to precision of our numerics, are both shown by solid blue. The curves show the DC conductivities obtained from the analytic results of Appendix~\protect\ref{app:DCderivation}, and the dots are the numerical results for optical conductivities extrapolated to $\omega = 0$. For both plots, the black dashed curve shows the DC conductivity of the homogenous phase (also at $\mu = 4$) for comparison.}
  \label{fig:DC_unpinned}
\end{figure}

Having obtained the sliding speeds, we can compute the DC conductivities as described in Appendix \ref{app:DCderivation}. Fig.~\ref{fig:DC_unpinned} shows the spatially averaged conductivities as functions of magnetic field.  The curves show the values computed with \eqref{finalDCsigmaxx}, \eqref{finalDCsigmaxy}, \eqref{finalDCsigmayy}, and \eqref{finalDCsigmayx}, while the dots are the $\omega \to 0$ limit of the optical conductivity computed in Sec.~\ref{sec:stripesAC}.

The spatially-averaged longitudinal conductivity across the stripes $\langle \sigma_{xx}^{DC} \rangle$ decreases with $b$,  in line with the decreasing sliding speed $\hat v_{x}$.  This current is largely carried by the  stripes sliding, as is evident from the dependence of $\langle \sigma_{xx}^{DC} \rangle$ on $\hat v_{x}$ in \eqref{finalDCsigmaxx}.   As was observed in \cite{Jokela:2016xuy} at $b=0$, the longitudinal conductivity parallel to the stripes $\langle \sigma_{yy}^{DC} \rangle$ is very nearly equal to $\langle \sigma_{xx}^{DC} \rangle$, in spite of the broken rotational symmetry in the striped phase.  Evidently, this not an artifact of zero magnetic field, as the near equality extends to nonzero $b$.  The close similarity is even more surprising considering that $\langle \sigma_{yy}^{DC} \rangle$ depends not on $\hat v_{x}$, but only on $\hat v_{y}$, which is increasing with $b$.   

The averaged Hall conductivity $\langle \sigma_{xy}^{DC} \rangle$, given by \eqref{finalDCsigmaxy}, depends linearly on $\hat v_y$, at least for small $b$.  This relationship is evident comparing Fig.~\ref{fig:DC_unpinned} (right) and Fig.~\ref{fig:vs}; the growth of $\langle \sigma_{xy}^{DC} \rangle$ closely matches that of $\hat v_y$  and implies that Hall sliding is largely responsible for the Hall current.

The other Hall conductivity $\langle \sigma_{yx}^{DC} \rangle$ is numerically equal to $\langle \sigma_{xy}^{DC} \rangle$ up to a sign, as shown in Fig.~\ref{fig:DC_unpinned} (right); the curves for  $-\langle \sigma_{yx}^{DC} \rangle$ and $\langle \sigma_{xy}^{DC} \rangle$ overlap.   The analytic expressions \eqref{finalDCsigmaxy} and \eqref{finalDCsigmayx}, however, do not appear to be obviously the same.\footnote{As we will see in Sec. \ref{sec:DCpinned}, once the stripes are pinned the analytic expressions for $-\langle \sigma_{yx}^{DC} \rangle$ and $\langle \sigma_{xy}^{DC} \rangle$ do become identical.} 
  This situation is like the near equality of the longitudinal conductivities $\langle \sigma_{xx}^{DC} \rangle$ and $\langle \sigma_{yy}^{DC} \rangle$ despite the clear asymmetry of the striped phase between the $x$ and $y$ directions.  In the Hall case, however, the equality between $-\langle \sigma_{yx}^{DC} \rangle$ and $\langle \sigma_{xy}^{DC}\rangle$ is exact, at least up to the precision of our numerics.

Comparing with the homogenous phase, the conductivity in the striped phase is higher at the same chemical potential.   This is largely because, for a given value of $\mu$, the striped phase has approximately 1 percent larger charge density $d$ \cite{Jokela:2014dba}.

\subsubsection{Inverted semi-circle law for stripes}
\label{sec:semi-circle_stripes}
In the homogenous phase, we saw above in Sec.~\ref{sec:semi-circle_homogenous} that the DC conductivities trace out a semi-circle in the Hall conductivity-longitudinal conductivity plane.  This relation is approximate for finite charge but becomes exact in the large $d$ limit.  However, at large $d$ the homogenous phase has a modulated instability and develops stripes.  

We find that the semi-circle law persists in the striped phase in a modified form for the spatially averaged conductivities.  Plotting the numerical values of $\langle \sigma_{xy}^{DC}\rangle$ against $\langle \sigma_{yy}^{DC}\rangle$ for various values of $b$ produces a portion of a semi-circle, as shown in Fig. \ref{fig:semicircle}.  At finite $\mu$, the striped phase does not exist beyond a maximum $b$, so the part of the semi-circle extending toward the origin is missing.

Now that the system is no longer isotropic, there is an ambiguity in the longitudinal conductivity.  We somewhat arbitrarily chose to plot $\langle \sigma_{yy}^{DC}\rangle$ rather than $\langle \sigma_{xx}^{DC}\rangle$.   However, because $\langle \sigma_{xx}^{DC}\rangle$ and $\langle \sigma_{yy}^{DC}\rangle$ are approximately equal and have similar $b$ dependence, this ambiguity is relatively minor.  If we had used $\langle \sigma_{xx}^{DC}\rangle$ instead, the semi-circle radius would be smaller but qualitatively the same.

\subsection{DC conductivity: pinned stripes}
\label{sec:DCpinned}

When either a magnetic or an ionic lattice is added to the inhomogeneous phase, $\alpha_b \ne 0$ or $\alpha_\mu \ne 0$,  the stripes become pinned and no longer slide.\footnote{As noted in \cite{Jokela:2017ltu}, the discontinuous change in behavior from $\alpha = 0$ to $\alpha > 0$ is an artifact of computing the linear response to an infinitesimal electric field. In the full, nonlinear conductivity we expect to see a de-pinning transition at some finite electric field.} Because the pinning prevents the stripes from sliding, it effectively distinguishes the portion of the current carried by the stripes from the portion which is not.

To determine the pinned DC conductivity, we simply set the sliding speed $v_s = 0$, as was described in \cite{Jokela:2017ltu} for vanishing magnetic field.  The (averaged) conductivities at nonzero $b$ can be read off from the 
formulas~\eqref{finalDCsigmaxx},  \eqref{finalDCsigmayy}, \eqref{finalDCsigmaxy},  and~\eqref{finalDCsigmayx}:
\begin{align}
\label{eq:sigmaDCxxpinned}
 \langle\sigma_{xx}^\mathrm{DC}\rangle &=  \langle\hat \sigma^{-1} \rangle^{-1} & \\
 \langle\sigma_{yy}^\mathrm{DC}\rangle &= \left\langle \hat \sigma(1+z_0'^2+\psi_0'^2) + \frac{1}{\hat \sigma}\left(\sqrt{2}c(\psi_0) 
 - \hat \sigma a_{t,0} (b+a_{y,0}'(x))\right)^2 \right\rangle  & \nn \\
 &\phantom{==}-  \langle\hat \sigma^{-1} \rangle^{-1} \left(\sqrt{2} \langle c(\psi_0) \hat \sigma^{-1} \rangle - \langle a_{t,0} (b+a_{y,0}') \rangle \right)^2&\\
 \langle\sigma_{xy}^\mathrm{DC}\rangle &=- \langle\sigma_{yx}^\mathrm{DC}\rangle = \left[\sqrt{2} \langle c(\psi_0) \hat \sigma^{-1} \rangle - \langle a_{t,0} (b+a_{y,0}') \rangle \right]\langle\hat \sigma^{-1} \rangle^{-1}  \ . &
\label{eq:sigmaDCxypinned}
\end{align}
Note that the Hall conductivities $\langle \sigma_{xy}^{DC} \rangle$ and $-\langle \sigma_{yx}^{DC} \rangle$ are now exactly equal.

Interestingly, the conductivities satisfy
\be\label{eq:stripesemicircle}
 \langle\sigma_{xx}^\mathrm{DC}\rangle \langle\sigma_{yy}^\mathrm{DC}\rangle +  \langle\sigma_{xy}^\mathrm{DC}\rangle^2 = 
\langle\sigma_{xx}^\mathrm{DC}\rangle \sigma_{yy}^0 \ ,
\ee
where
\be
 \sigma_{yy}^0 = \left\langle \hat \sigma(1+z_0'^2+\psi_0'^2) + \frac{1}{\hat \sigma}\left(\sqrt{2}c(\psi_0) 
 - \hat \sigma a_{t,0} (b+a_{y,0}'(x))\right)^2 \right\rangle
\ee
is the term with parity-even averages in $ \langle\sigma_{yy}^\mathrm{DC}\rangle$.  The relation \eqref{eq:stripesemicircle} can be viewed as the analog of the semi-circle law, modified for the anisotropic striped phase; for an isotropic phase, $\langle\sigma_{xx}^\mathrm{DC}\rangle = \langle\sigma_{yy}^\mathrm{DC}\rangle$, and  \eqref{eq:stripesemicircle} would resemble a semi-circle.

\begin{figure}[!ht]
\center
 \includegraphics[width=0.50\textwidth]{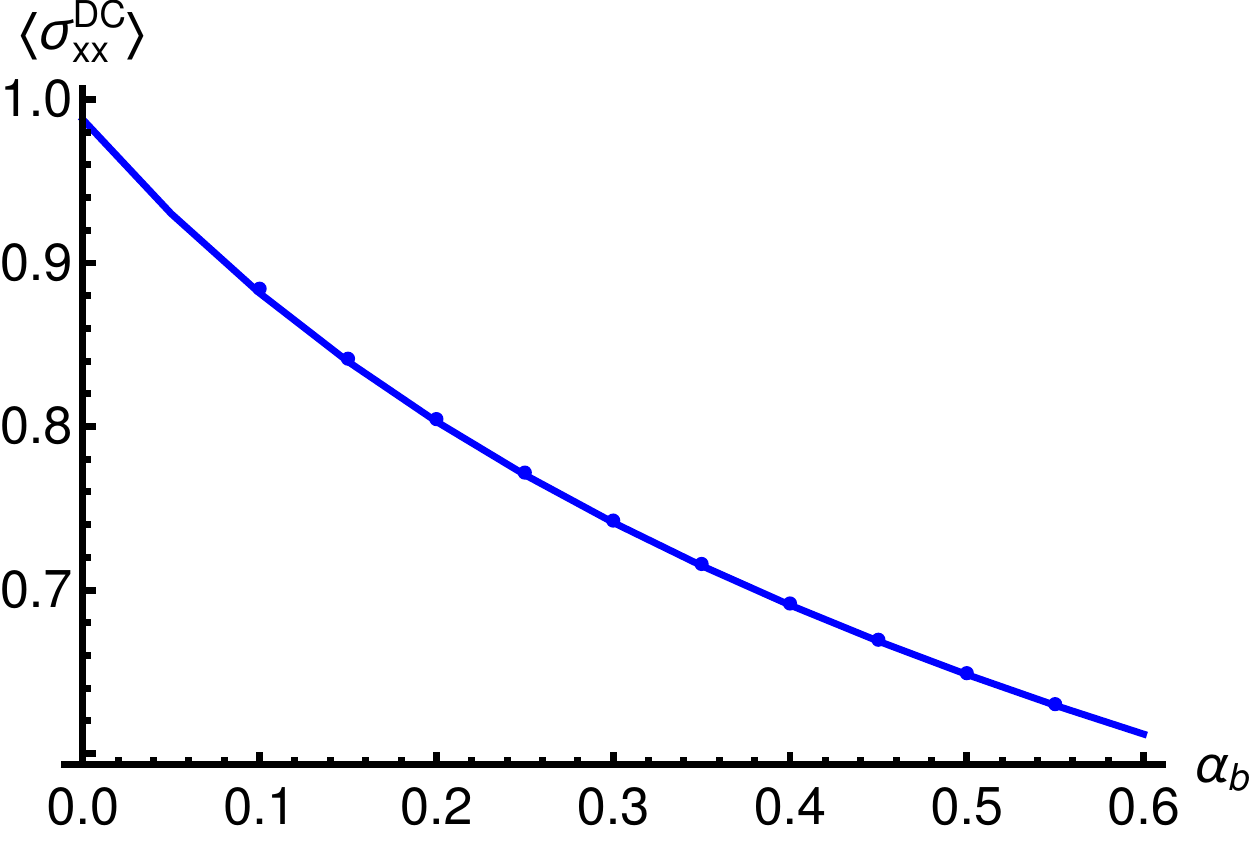}%
 \includegraphics[width=0.50\textwidth]{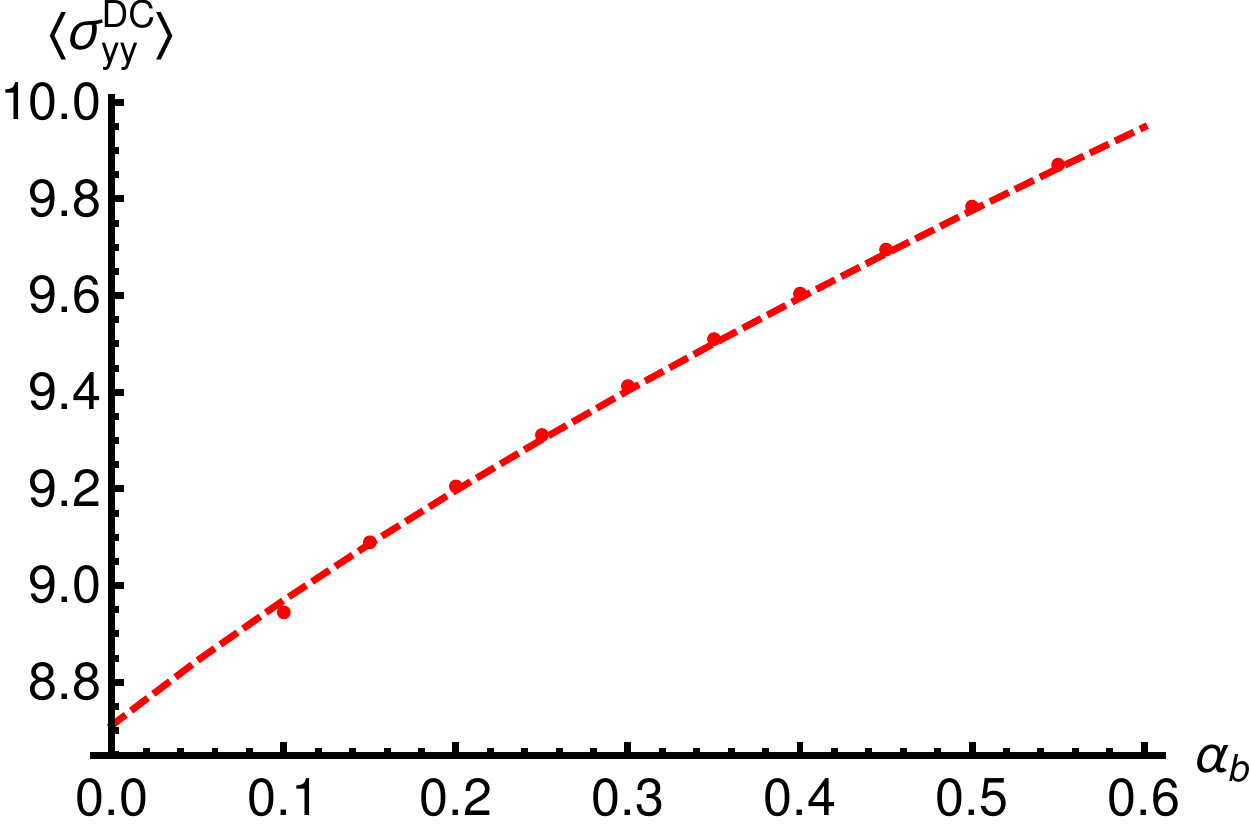}
 \includegraphics[width=0.50\textwidth]{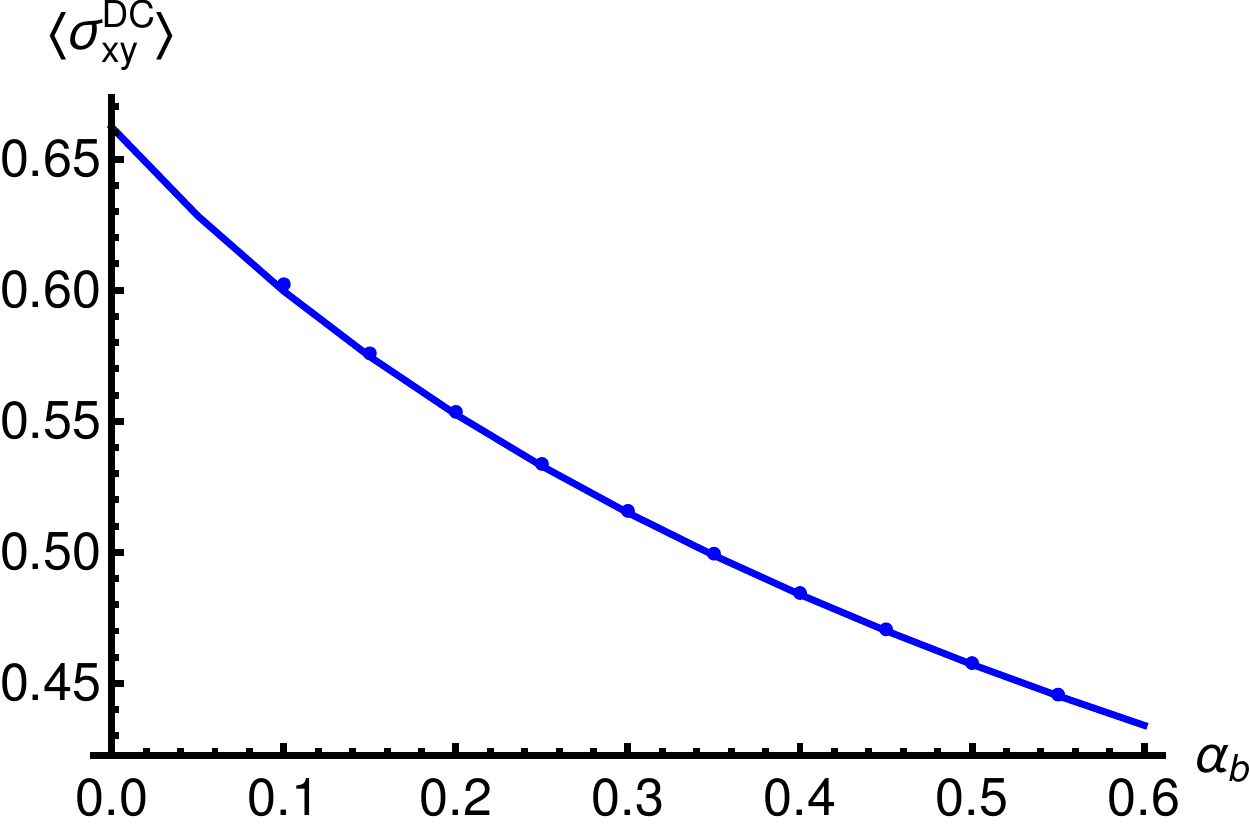}%
 \caption{The averaged pinned DC conductivities for magnetic lattice at $b=0.5$. The curves show the analytic results from Eqs.~\protect\eqref{eq:sigmaDCxxpinned}--\protect\eqref{eq:sigmaDCxypinned}, and the dots are are the optical conductivities extrapolated to $\omega = 0$.} 
  \label{fig:DC_magnetic}
\end{figure}

The DC conductivities at nonzero $b$, pinned by a magnetic lattice of varying strength $\alpha_b$ are shown in Fig.~\ref{fig:DC_magnetic}.  As expected, the currents across the stripes and the lattice are significantly reduced.  Compared with the unpinned results, shown in Fig.~\ref{fig:DC_unpinned}, the conductivities $\langle \sigma_{xx}^{DC} \rangle$ and $\langle \sigma_{xy}^{DC} \rangle$ are both an order of magnitude smaller and decrease with $\alpha_b$.   As was seen in \cite{Jokela:2017ltu}, the longitudinal conductivity parallel the stripes $\langle \sigma_{yy}^{DC} \rangle$ actually increases with the strength of the lattice.  However, the Hall conductivity $-\langle \sigma_{yx}^{DC} \rangle$, which is equal to $\langle \sigma_{xy}^{DC} \rangle$, decreases with $\alpha_b$.

\subsection{Optical Conductivity for stripes}
\label{sec:stripesAC}

We compute the optical conductivity of the striped phase as in \cite{Jokela:2016xuy}, but now with a nonzero magnetic field.  As in the homogeneous phase in Sec.~\ref{sec:optical_conductivity_homogeneous} above, we perform a linear fluctuation analysis, but with applied electric fields in both the $x$ and $y$ directions.  As was discussed above in Sec.~\ref{sec:DCunpinned}, the range of the magnetic field is limited because, at a critical $b$, there is a phase transition to the homogenous phase, and just beyond that, the striped phase no longer exists.

\begin{figure}[!ht]
\center
 \includegraphics[width=0.50\textwidth]{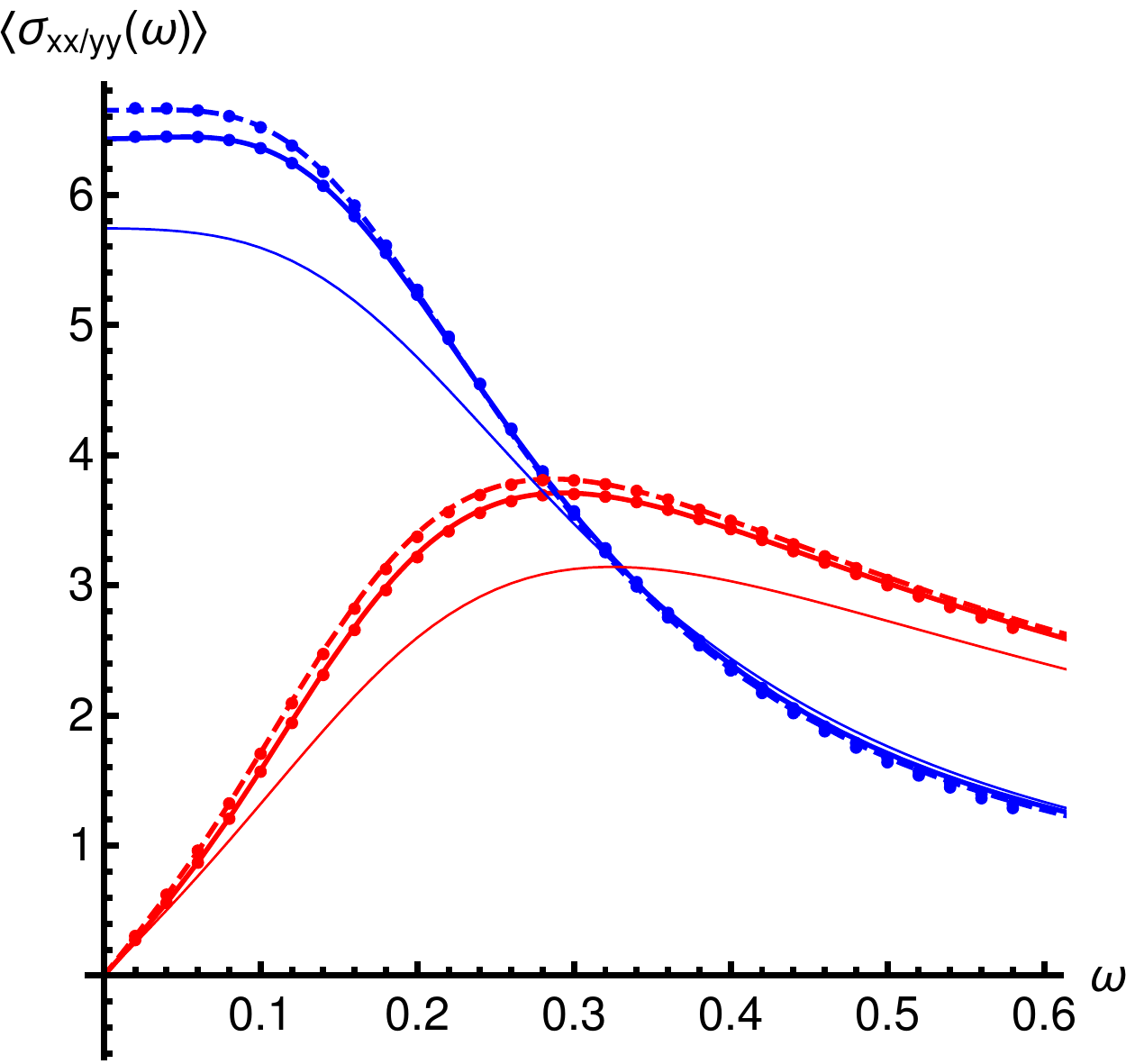}%
 \includegraphics[width=0.50\textwidth]{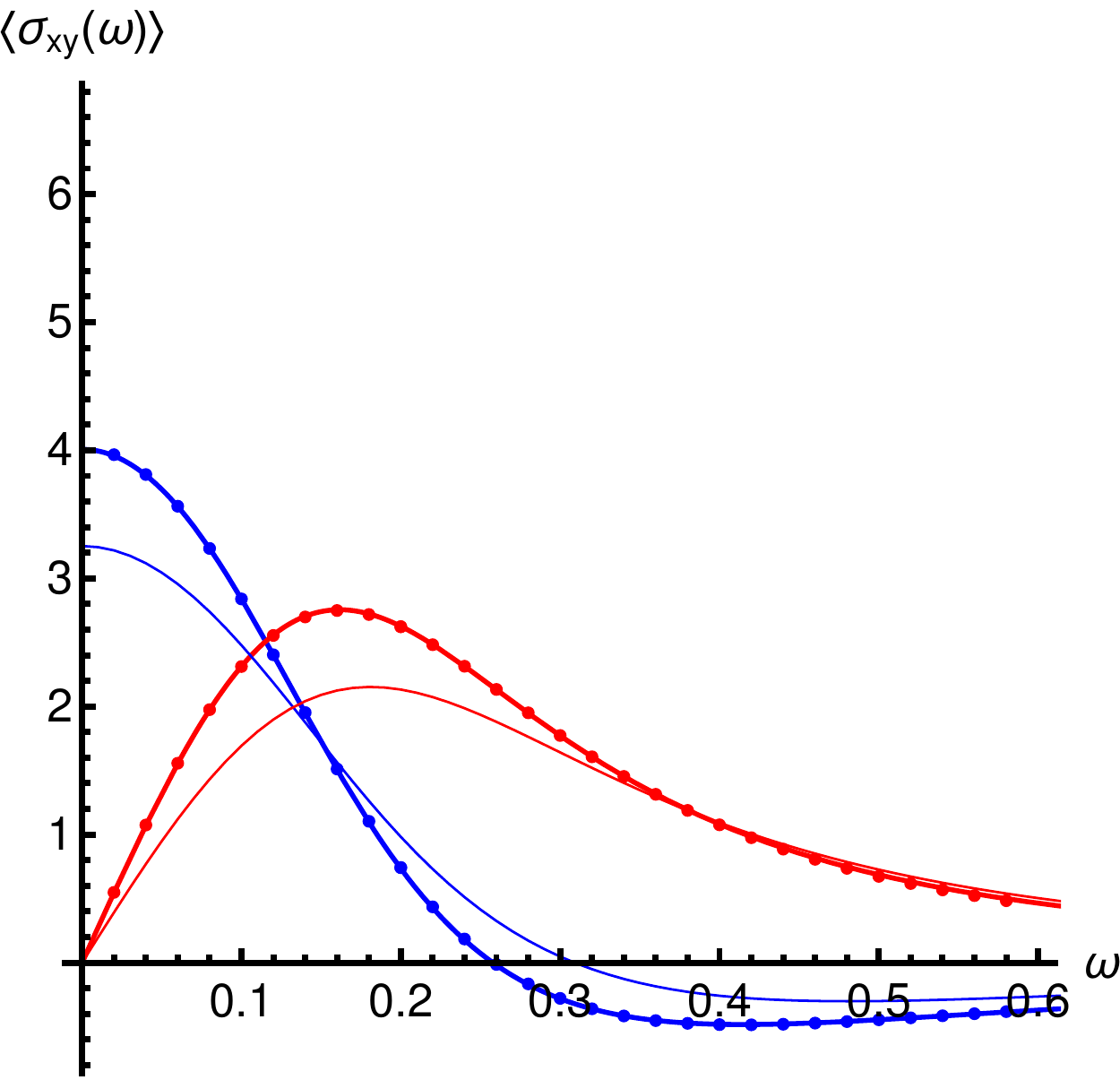}
 \caption{The spatially-averaged  longitudinal (left) and  Hall (right) optical conductivities of the striped phase, fit to analytic formulas at $b=0.5$ and $\mu=4$, with the real parts shown in blue and the imaginary parts in red. Dots are the numerically computed data, and the curves show the fits.  (left) The dashed curve is $\langle \sigma_{xx}\rangle$, and the solid is  $\langle \sigma_{yy}\rangle$.  (right) The solid curve is $\langle \sigma_{xy}\rangle$ which is numerically equal to $-\langle \sigma_{yx}\rangle$. In both plots, the thin, solid curves are the analogous conductivities of the homogenous phase, shown for comparison.} 
  \label{fig:unpinned_fit} 
\end{figure}

As a representative example, the spatially averaged conductivities for $b=0.5$ are shown in Fig.~\ref{fig:unpinned_fit}.   On the left, $\langle \sigma_{xx}(\omega)\rangle$ and $\langle \sigma_{yy}(\omega)\rangle$ continue to be just similar at nonzero frequency as was seen in see Sec.~\ref{sec:DCunpinned}) at $\omega = 0$, in spite of the spontaneously broken rotational symmetry.  On the right, $\langle \sigma_{xy}(\omega)\rangle$ is still equal to  $-\langle \sigma_{yx}(\omega)\rangle$, to the precision of our numerics. 

For comparison, the thin solid curves represent the analogous conductivities for the homogeneous phase at the same values of $\mu$ and $b$.  Again, the difference between the homogeneous and striped results is due primarily to the striped phase having a slightly larger charge density at a given chemical potential.

Our results once again match, to reasonable accuracy, the hydrodynamical model of \cite{Hartnoll:2007ih}.  Both longitudinal conductivities, $\sigma_{xx}$ and $\sigma_{yy}$, are fit to \eqref{Hartnoll_sigmaxx}, and the Hall conductivities, $\sigma_{xy}$ and $\sigma_{yx}$, are fit to \eqref{Hartnoll_sigmaxy}.  As in Sec.~\ref{sec:optical_conductivity_homogeneous}, the values of $\sigma_Q$ and $\rho$ are fixed by matching the zero-frequency limit with the DC conductivities obtained in Sec.~\ref{sec:DCunpinned}.  The remaining parameters to be fit by the data are $\gamma_v$, $\tau$, and $\kappa_\omega$.  We fit the parameters separately for each component of the conductivity.  In principle, $\gamma_v$, $\tau$, and $\kappa_\omega$ should be the same for each component, and the similarity of different curves indicates the degree of accuracy of the model.  For example, the proportionality constant $\kappa_\omega$ of the cyclotron frequency is approximately 0.25 for all three components, at least for $b$ sufficiently large,\footnote{As noted above in Sec.~\ref{sec:optical_conductivity_homogeneous}, because the scaling with $b$ was factored out, the fit becomes insensitive to $\kappa_\omega$ at small $b$ and the growth as $b\to 0$ is a numerical artifact of this.} which is roughly the same value found in the homogenous case (see Fig.~\ref{fig:fit_params_homog}).

\begin{figure}[!ht]
\center
 \includegraphics[width=0.50\textwidth]{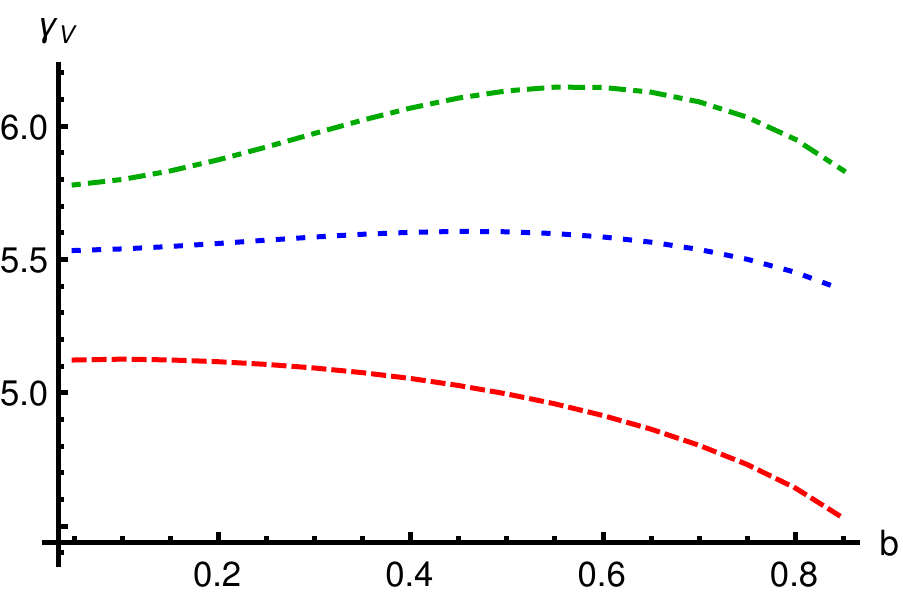}%
 \includegraphics[width=0.50\textwidth]{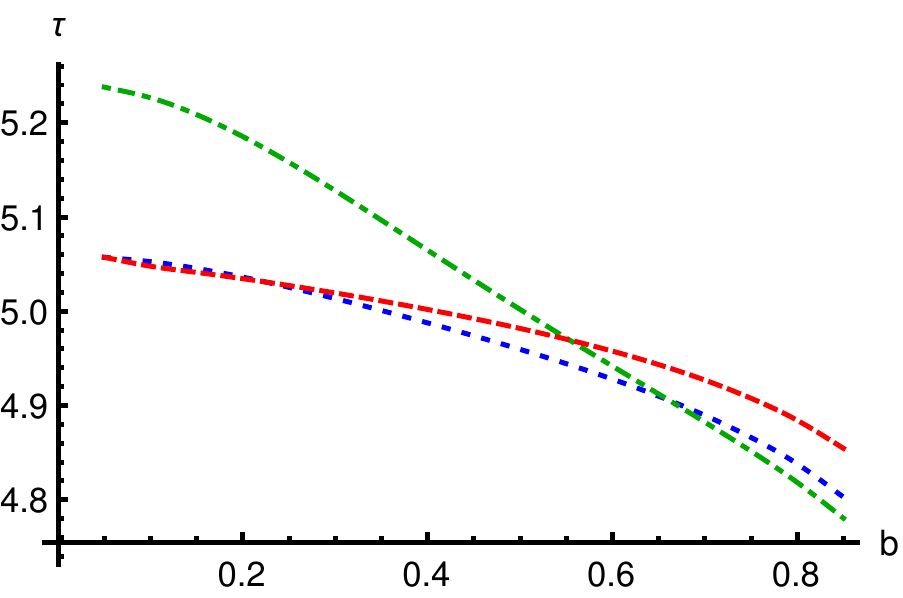}
 \includegraphics[width=0.50\textwidth]{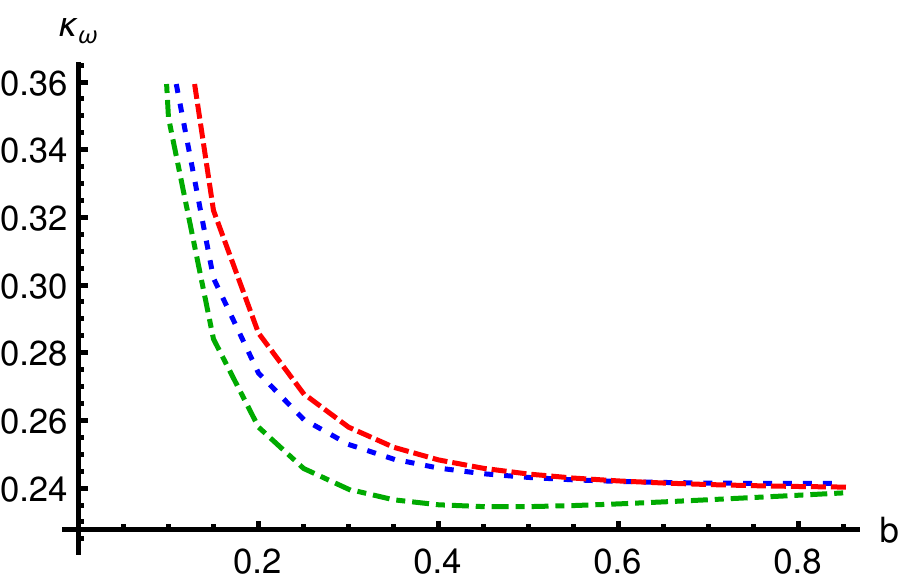}%
 \includegraphics[width=0.50\textwidth]{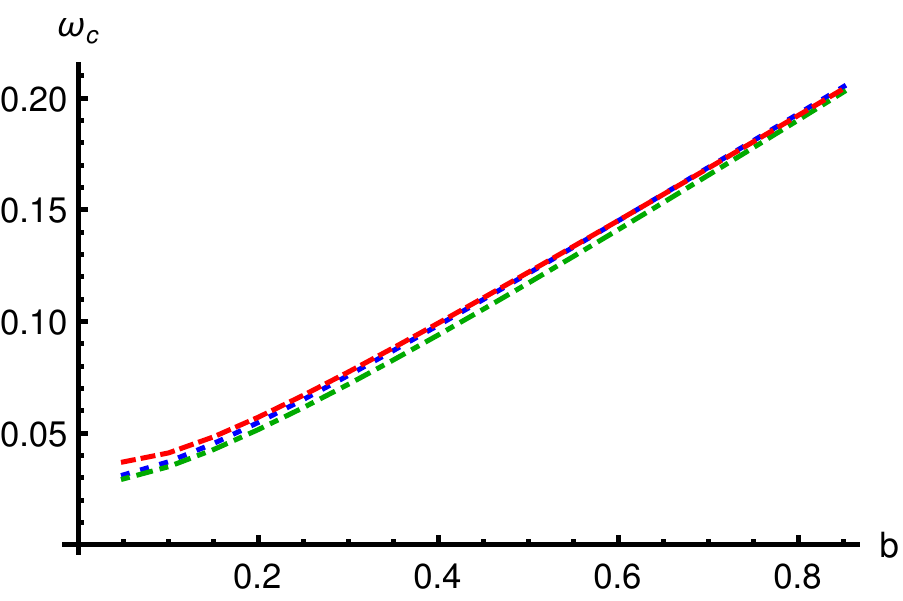}
 \caption{The fit parameters for averaged optical conductivities of the striped phase as functions of $b$, for $\mu = 4$. The dotted blue, dot-dashed green, and dashed red curves are fits to data for $\langle\sigma_{xx}\rangle$, $\langle\sigma_{yy}\rangle$, and $\langle\sigma_{xy}\rangle$, respectively.}  
  \label{fig:fit_params}
\end{figure}

As was seen in Sec.~\ref{sec:optical_conductivity_homogeneous} for the homogeneous phase, the peak in the longitudinal conductivity is at the cyclotron frequency $\omega_c$.  As can be seen in Fig.~\ref{fig:fit_params}, $\omega_c$ is again proportional to $b$ to good accuracy.  In Fig.~\ref{fig:unpinned_fit}, the relatively small magnetic field produces a cyclotron peak close to zero frequency which, as a result, is less visible in the plot.

A notable feature observed in \cite{Jokela:2016xuy} was a delta function at $\omega = 0$ in the Hall conductivity $\sigma_{yx}$. A perturbative electric field in the $x$ direction causes the stripes to slide with a speed proportional to $v_x$, carrying along the persistent transverse current and changing the local current in the $y$ direction by a finite, non-perturbative amount equal to the divergence of the background current $J_y'(x)$. This divergence is spatially modulated and is eliminated by spatial averaging.  Numerically, the presence of the delta function was observed in the presence of a pole in the imaginary part of $\sigma_{yx}(x)$:
\be
 \sigma_{yx}(x) \sim - \frac{iK_{yx}(x)}{\omega} \ ,
\ee
where the weight of the pole is $K_{yx}(x) = \hat v_x J_y'(x)$

With the addition of a background magnetic field, an electric field applied in the $y$ direction also causes the stripes to slide. This Hall sliding also carries along the persistent transverse current, changing the local current in the $y$ direction and resulting in a spatially modulated delta function in $\sigma_{yy}(x)$:
\be
 \sigma_{yy}(x) \sim - \frac{iK_{yy}(x)}{\omega} \ ,
\ee
where the weight of the pole is proportional to the sliding speed $\hat v_y$,  $K_{yy}(x) = \hat v_y J_y'(x)$.

\begin{figure}[!ht]
\center
  \includegraphics[width=0.50\textwidth]{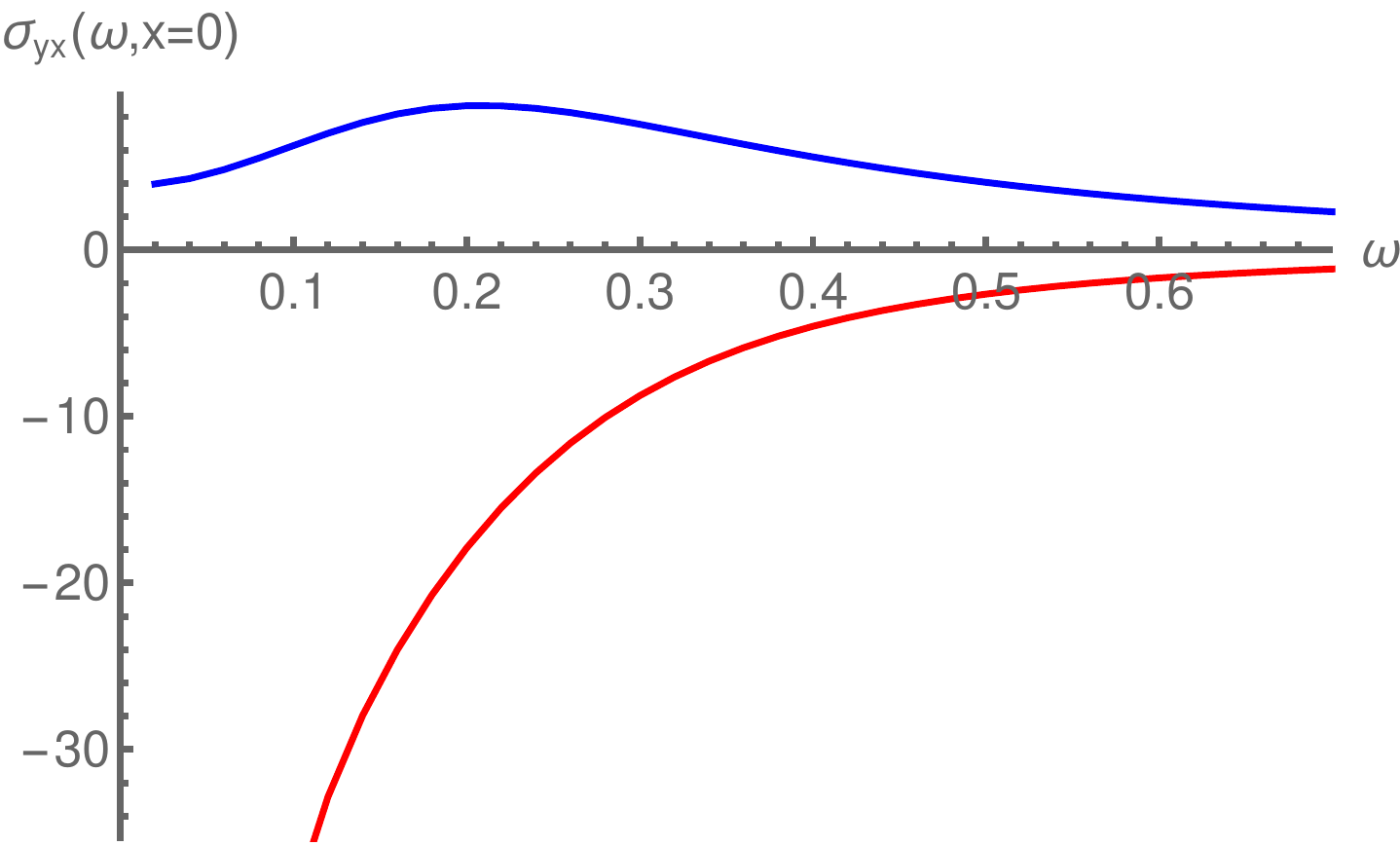}%
 \includegraphics[width=0.50\textwidth]{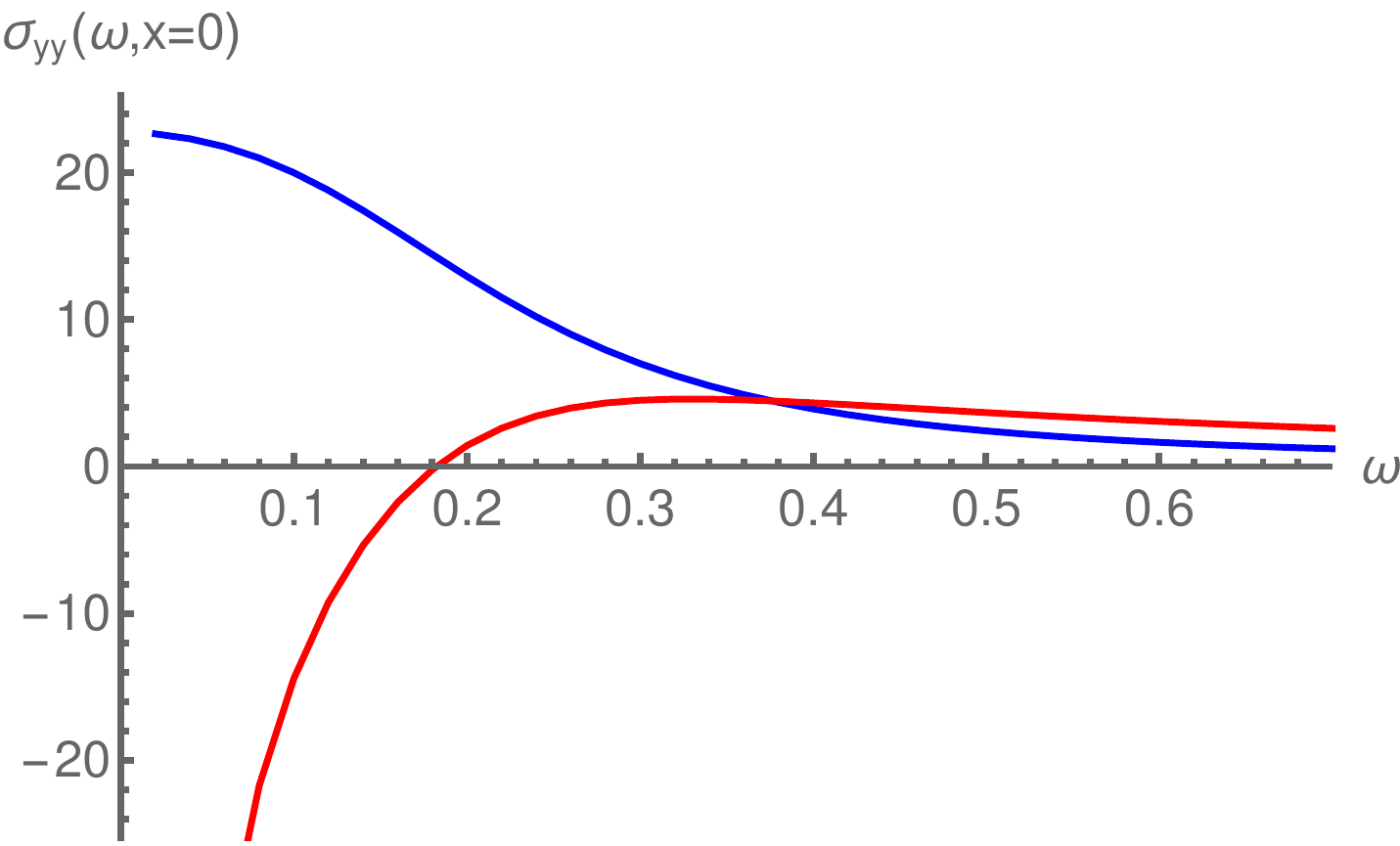}
 \caption{
 The components of the conductivity (left) $\sigma_{yx}$ and (right) $\sigma_{yy}$ at $b=0.5$ featuring spatially modulated delta functions at zero frequency, shown at $x=0$. Real parts are blue, and imaginary parts are red.  Although the delta function in the real part is not accessible numerically, the corresponding pole in the imaginary part is visible.  }   
  \label{fig:magnsigma}
\end{figure}

\begin{figure}[!ht]
\center
 \includegraphics[width=0.50\textwidth]{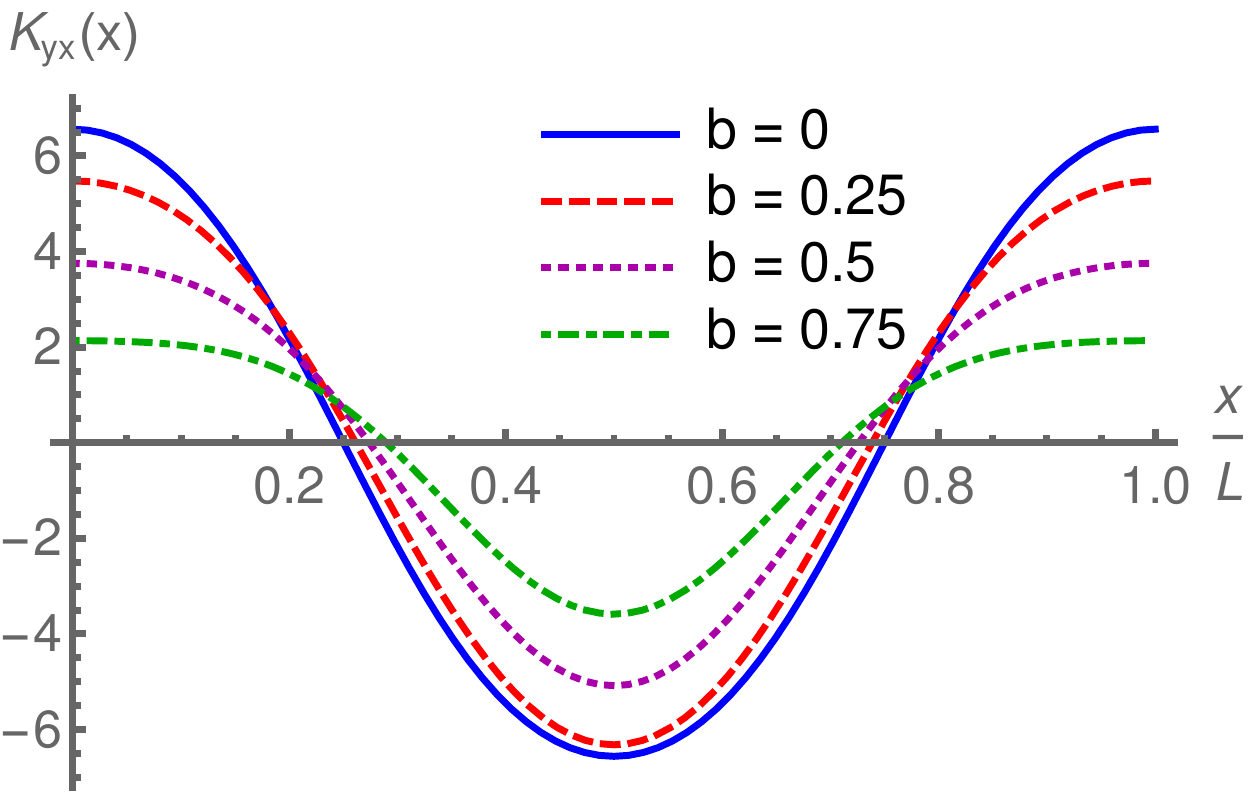}%
 \includegraphics[width=0.50\textwidth]{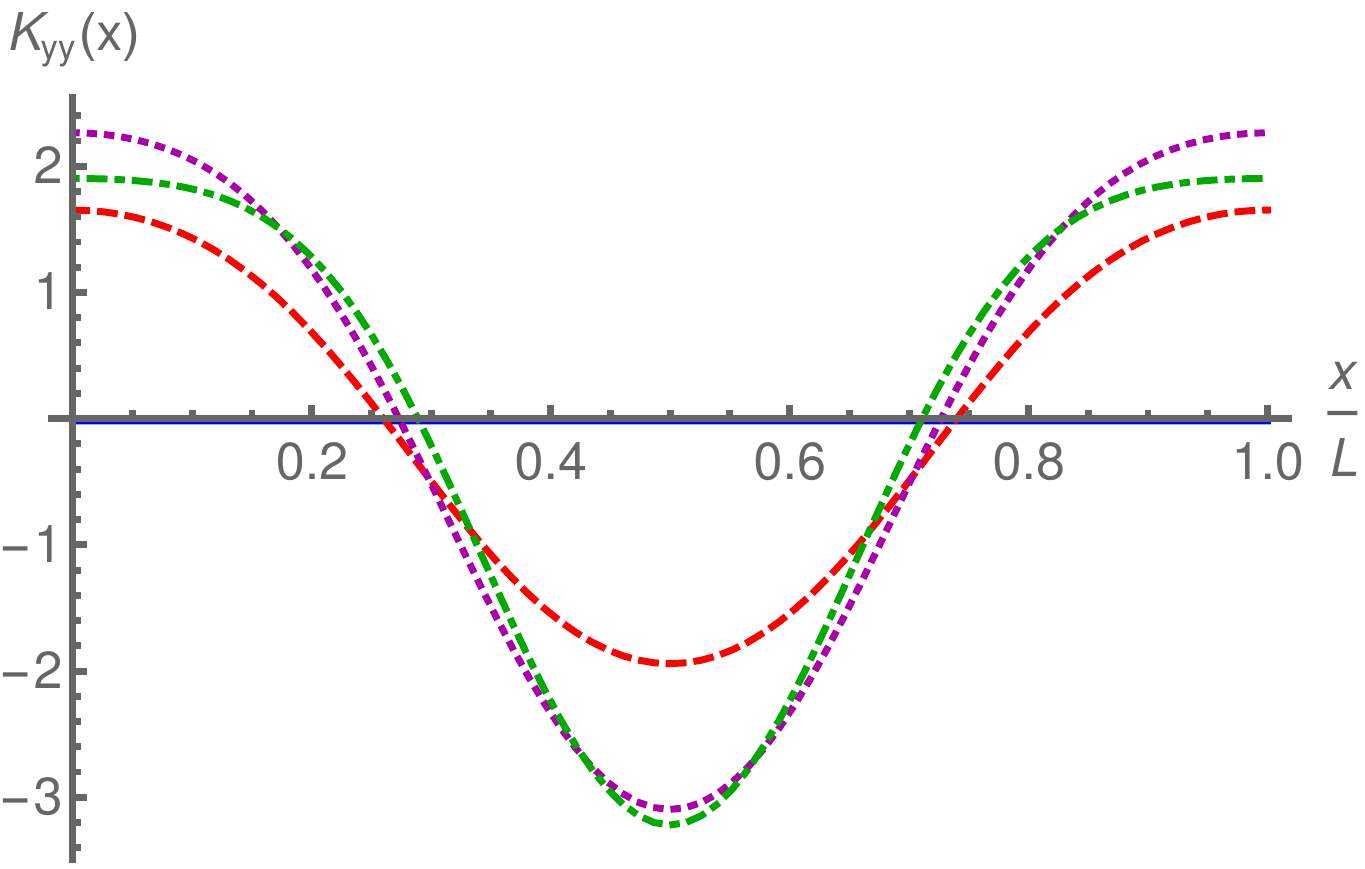}
 \includegraphics[width=0.50\textwidth]{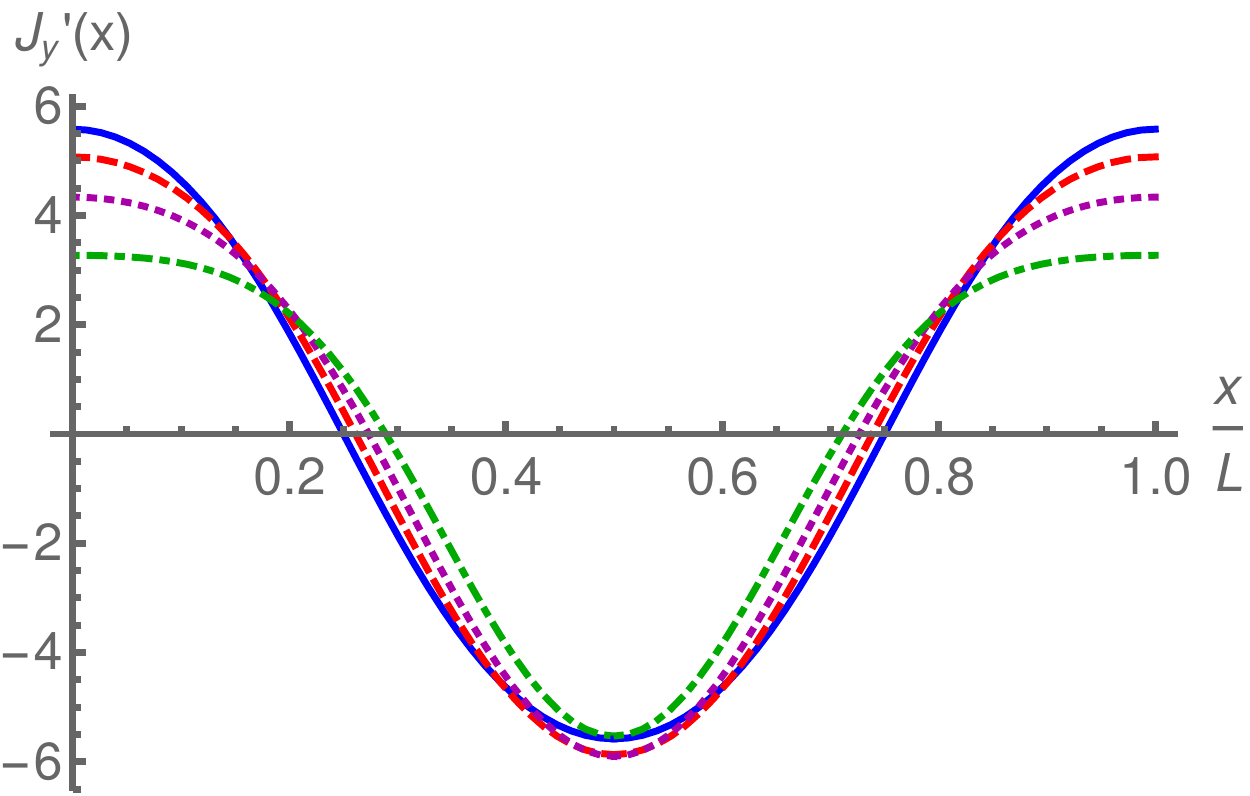}
 \caption{(Top row) 
 The $x$-dependence of the weights of the delta functions: 
 (Left) The weight $K_{yx}(x)$ in the optical conductivity $\sigma_{yx}(\omega)$;  
 (Right) The weight $K_{yy}(x)$ in the optical conductivity $\sigma_{yy}(\omega)$.  (Bottom) The derivative of the background current $J_y'(x)$. 
 The blue, red dashed, magenta dotted, and green dot-dashed curves  are for  $b=0$, $0.25$, $0.5$, and $0.75$, respectively.}   
  \label{fig:Kfactors}
\end{figure}

The conductivities $\sigma_{yx}$ and $ \sigma_{yy}$ at $x=0$ (i.e. not spatially averaged) are shown in 
Fig. \ref{fig:magnsigma}. The imaginary parts, shown as solid red curves, clearly exhibit a pole at $\omega=0$.  The weights  $K_{yx}(x)$ and $K_{yy}(x)$, for various values of $b$ are shown in the top 
row of 
Fig.~\ref{fig:Kfactors}.  At zero $b$, $K_{yy}(x) = 0$, because the Hall sliding only occurs at nonzero magnetic field.  As the magnetic field is increased, the amplitude of the modulated pole $|K_{yx}|$ decreases because the sliding speed $v_x$ decreases (see Fig.~\ref{fig:vs}).  The Hall sliding speed $v_y$, however, increases with $b$, causing $|K_{yy}|$ also to increase.  In addition, the $x$ dependence of $J_y'(x)$ becomes asymmetric 
(see the bottom plot in Fig.~\ref{fig:Kfactors}), 
making $K_{yx}(x)$ and $K_{yy}(x)$ similarly  distorted.

\subsection{Optical Conductivity for pinned stripes}

\begin{figure}[!ht]
\center
 \includegraphics[width=0.50\textwidth]{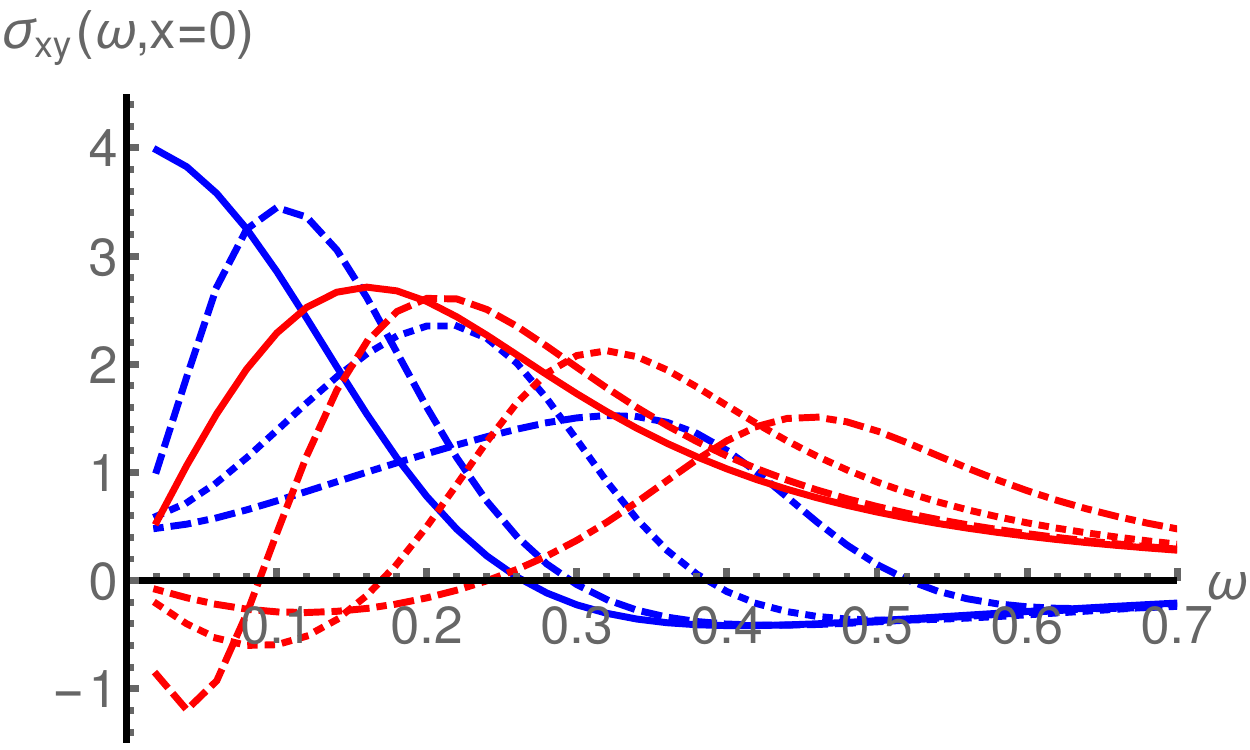}%
 \includegraphics[width=0.50\textwidth]{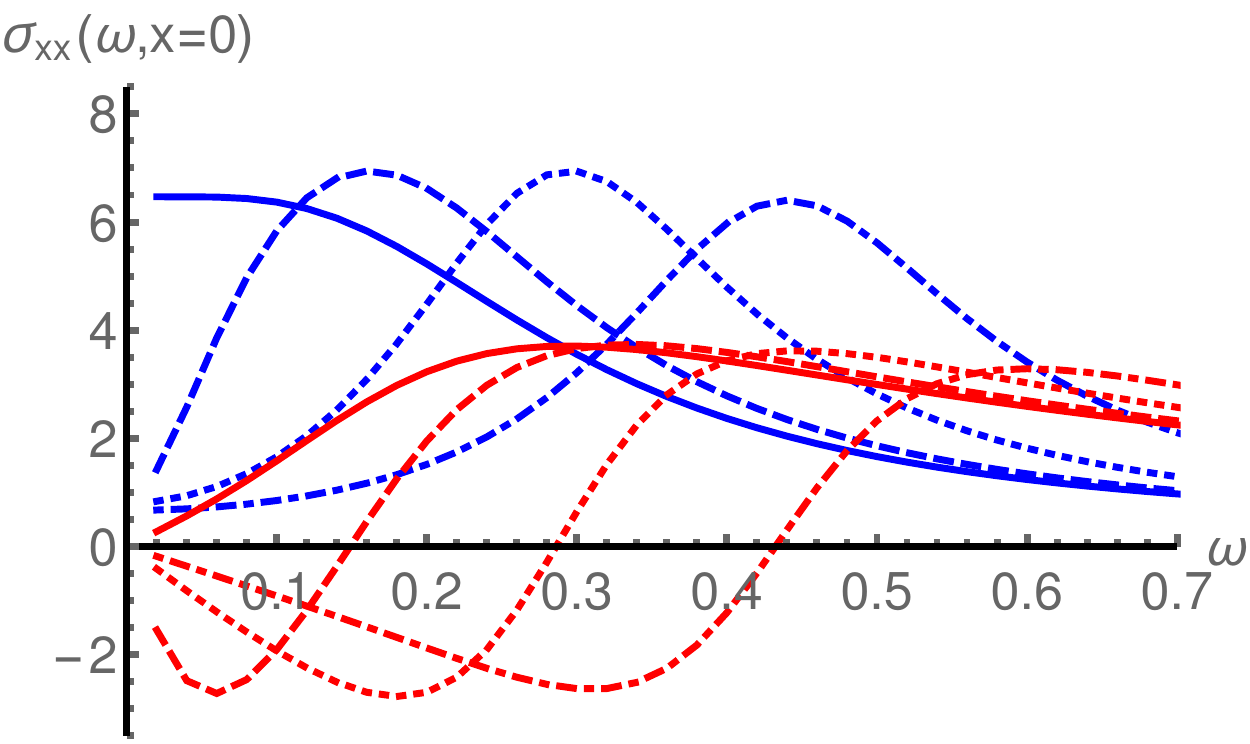}
 \caption{The pinning effect due to the magnetic lattice on the components of the conductivity that do not feature delta peaks, at $b=0.5$ and $x=0$. The blue (red) curves are real (imaginary) parts, and the solid, dashed, dotted, and dot-dashed curves are for $\alpha_b = 0$, $0.1$, $0.25$, and $0.5$, respectively. Left: the behavior of $\sigma_{xy}$ at $x=0$.  Right: the behavior of $\sigma_{xx}$ at $x=0$.}   
  \label{fig:pinnednodeltapeaks}
\end{figure}

\begin{figure}[!ht]
\center
 \includegraphics[width=0.50\textwidth]{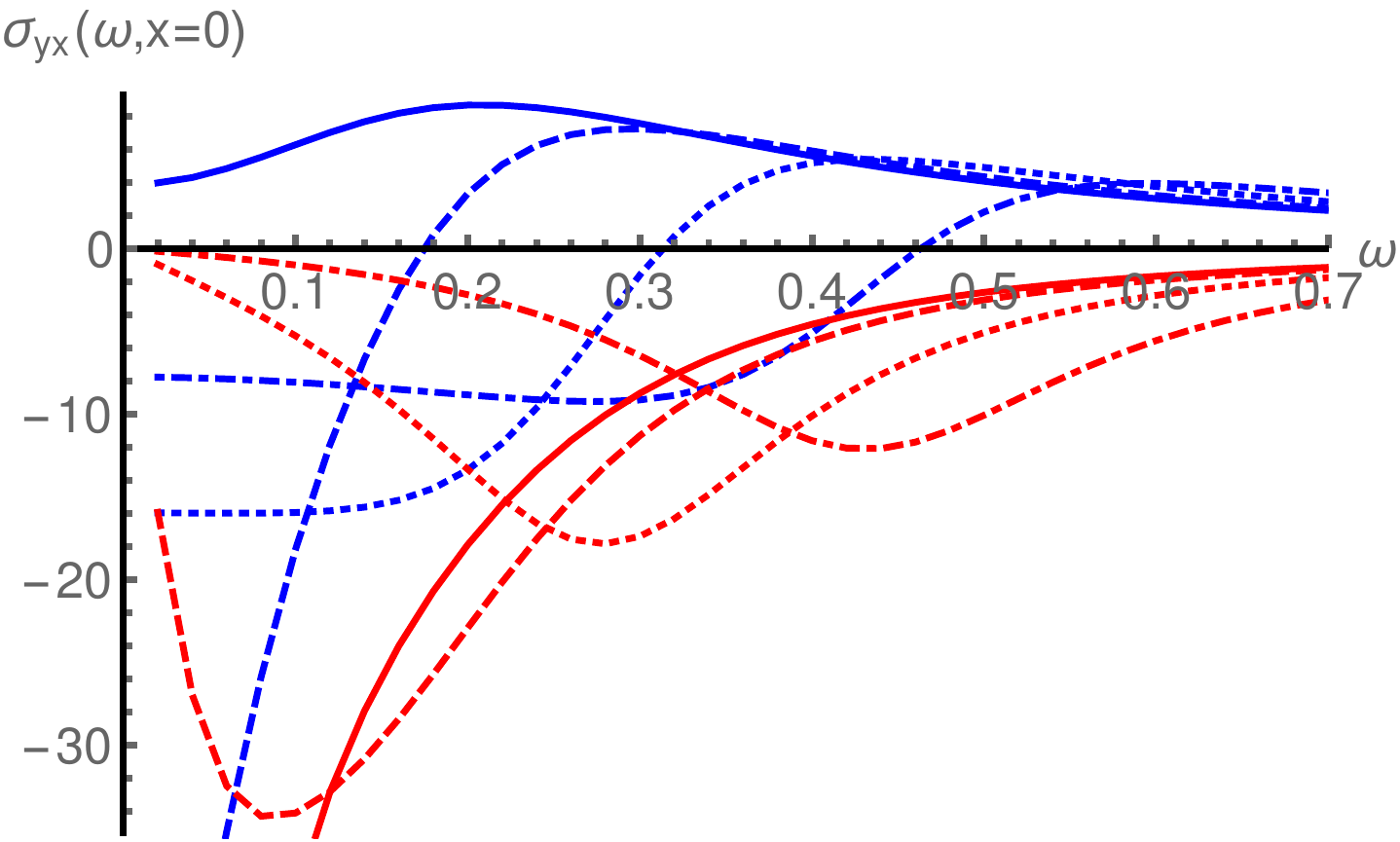}%
 \includegraphics[width=0.50\textwidth]{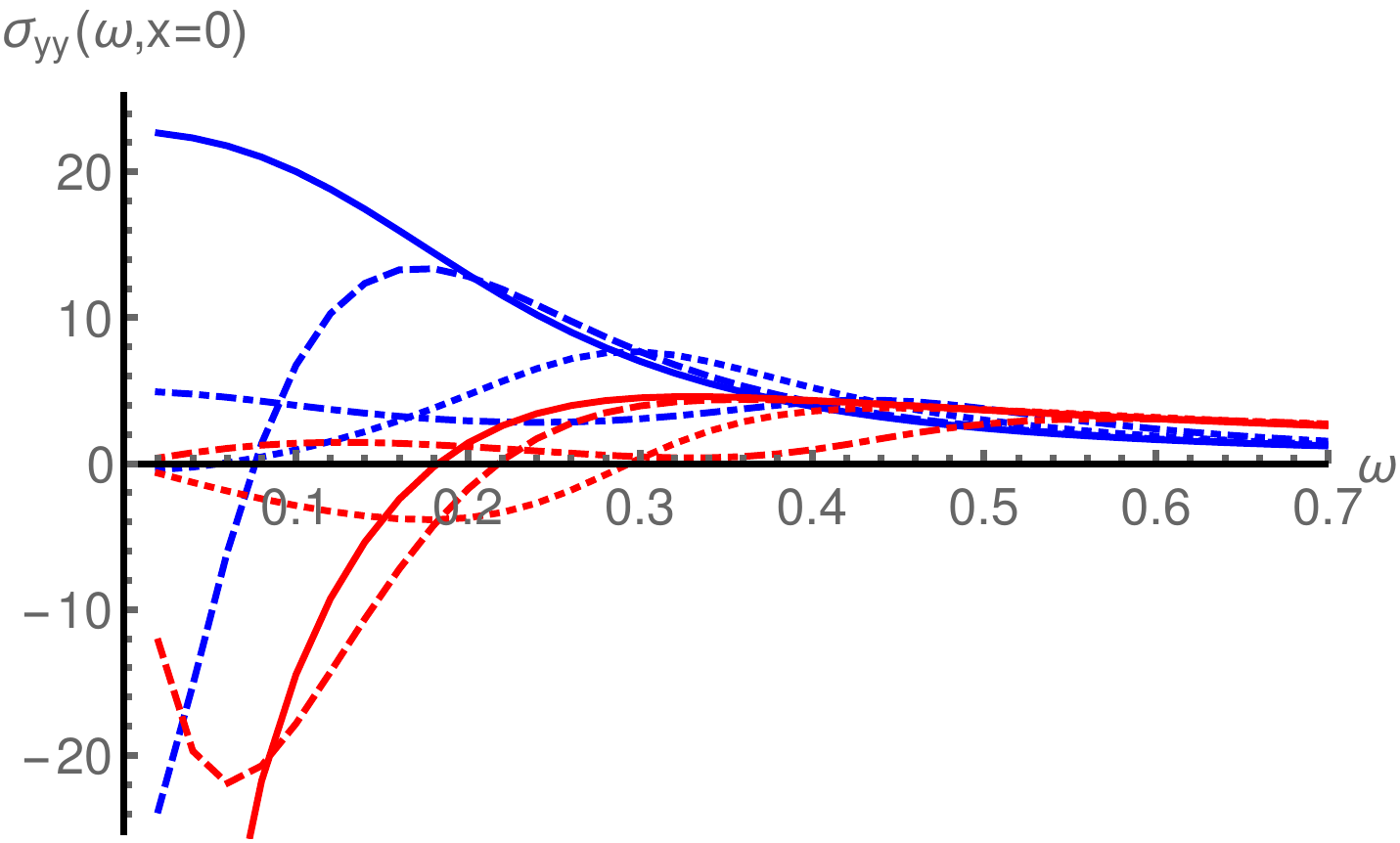}
 \caption{The pinning effect due to the magnetic lattice on the components of the conductivity that do  feature delta peaks, at $b=0.5$ and $x=0$. The blue (red) curves are real (imaginary) parts, and the solid, dashed, dotted, and dot-dashed curves are for $\alpha_b = 0$, $0.1$, $0.25$, and $0.5$, respectively. Left: the behavior of $\sigma_{yx}$ at $x=0$.  Right: the behavior of $\sigma_{yy}$ at $x=0$.}   
  \label{fig:pinneddeltapeaks}
\end{figure}

With the addition of a lattice, we can investigate the interplay between the effects of the constant magnetic field and the pinning effect of the lattice.  We numerically computed the optical conductivity, as above in Sec.~\ref{sec:stripesAC}, but with the addition of a background magnetic lattice of strength $\alpha_b$. To illustrate a representative example, the optical conductivities at fixed $b = 0.5$, with several values of $\alpha_b$, are shown in Figs. \ref{fig:pinnednodeltapeaks} and \ref{fig:pinneddeltapeaks}. 

When $\alpha_b > 0$, the pinned stripes no longer slide, as we saw at $b=0$ in \cite{Jokela:2017ltu}, which can be observed in the conductivity in two ways.  As was found in Sec.~\ref{sec:DCpinned}, the DC conductivity drops by an order of magnitude when the lattice is turned on.  In addition, the modulated delta function resulting from the sliding of the persistent current disappears.  As shown in Fig.~\ref{fig:pinneddeltapeaks}, the associated pole at $\omega = 0$ in the imaginary parts of $\sigma_{yx}$ and $\sigma_{yy}$ only exists for $\alpha_b = 0$.  At nonzero $\alpha_b$, the imaginary parts of $\sigma_{yx}$ and $\sigma_{yy}$ are suppressed at zero frequency.

In the absence of a constant background magnetic field, we found in \cite{Jokela:2017ltu} that pinning causes the Drude peak in $\sigma_{xx}$ to move to nonzero frequency.  This pinning pole is due to the damped harmonic oscillation of the stripes in the lattice potential.   

Here, we find that the conductivity in a constant background magnetic field closely resembles the results found in \cite{Jokela:2017ltu}.   For the unpinned stripes, the cyclotron pole at $b=0.5$ is located at $\omega_c \approx 0.12$, and when pinning is turned on, the peak in the real part shifts to higher frequencies just as in \cite{Jokela:2017ltu}.  The  peak also broadens and shrinks, as the corresponding pole moves further off the real axis.  

These results are in contrast to the two separate poles observed in \cite{Song:2019rnf}.  In \cite{Song:2019rnf}, the cyclotron pole observed in the absence of pinning and the pinning pole seen at zero magnetic field remained distinct when the two effects were combined.  The disparity could be due to differences in the holographic construction and implementation of the pinning.  Alternatively, it could be that, in the D3-D7' model, the cyclotron and pinning poles are not well separated.  In particular, the strength of the magnetic field in the striped phase is limited; at large magnetic field, the system is in the homogenous phase.\footnote{For the small values of $\alpha_b \lesssim 1$ used here, the pinning pole is found at $\omega \lesssim 1$.  For a cyclotron pole with $\omega_c > 1$, a magnetic field with $b$ of order 10 would be needed.  However, for the charge density $\mu =4$, the magnetic field in striped phase can be at most 0.95.}

In \cite{Jokela:2017ltu}, the conductivity at $b=0$ was accurately fit by a Drude-Lorentz model.  The Lorentzian described the damped, driven oscillation of the pinned stripes, while the Drude term captured the residual metallic current across the stripes.  As seen above, the conductivity at nonzero $b$ is fit by the hydrodynamical model \eqref{Hartnoll_sigmaxx} and \eqref{Hartnoll_sigmaxy}. When pinned, these expressions are expected to be  modified to include an additional pinning peak \cite{Delacretaz:2019wzh} (see also \cite{Chen:2005}). However, our results at nonzero $b$ also appear to be well fit by the Drude-Lorentz model used at $b=0$ in \cite{Jokela:2017ltu}, making it challenging to differentiate between the \eqref{Hartnoll_sigmaxx} and \eqref{Hartnoll_sigmaxy} and the results of \cite{Delacretaz:2019wzh}. 


\section{Discussion and open problems}\label{sec:Discussion}

In this paper, we investigated the electrical conductivity of a strongly-coupled (2+1)-di\-men\-sion\-al fermionic fluid in a background magnetic field.
The hydrodynamic model of \cite{Hartnoll:2007ih} seems to closely match our results.  In particular, we observe a cyclotron peak in the conductivity at a frequency proportional to the magnetic field.  We observed Hall sliding due to an electric field along the stripes, which is an expected generalization of the sliding due to an electric field across the stripes seen in \cite{Jokela:2016xuy}.  The sliding stripes, in both the longitudinal and Hall cases, carry significant amounts of charge and remain important contributors to the conductivity.

We encountered a number of puzzling phenomenological observations for which we lack a fundamental theoretical explanation.
One notable discrepancy with similar holographic studies is the lack of separate cyclotron and pinning poles, as was found in \cite{Delacretaz:2019wzh, Song:2019rnf}.  Although differences in the models could be the reason, the single pole we find may be an artifact of our necessarily limited parameter space.  For example, in \cite{Baggioli:2020edn}, the magneto-conductivity exhibited a single magnetophonon pole while the cyclotron pole was absent, having been pushed to large frequency.  A more thorough investigation of the QNM spectrum of our model would clarify this issue.

The symmetry or approximate symmetry between conductivities across and along the stripes, in spite of the broken rotational symmetry of the stripe phase and also the broken parity due to the magnetic field, remains unexplained.  The longitudinal conductivities $\sigma_{xx}$ and $\sigma_{yy}$ were found to be approximately equal in \cite{Jokela:2016xuy} and continue to be so in the presence of a magnetic field.  Perhaps more surprisingly, we found that the spatially averaged Hall conductivities $\langle \sigma_{xy}\rangle$ and $-\langle \sigma_{yx}\rangle$ resulting from the applied magnetic field appear to be exactly equal.  This symmetry is especially puzzling considering the large role the sliding plays in transporting charge in the $x$ direction but not the $y$ direction.

Although the novel semi-circle law in the DC conductivity is described analytically, at least in certain limit, we lack an understanding of its theoretical origin. The behavior is distinct enough from the usual semi-circle law to suggest a rather different physical explanation.

Having obtained surprisingly good results using a hydrodynamical description, it would be an interesting exercise to understand it better in the probe brane case as studied here.\footnote{We note that the theoretical foundation of the hydrodynamic framework of \cite{Hartnoll:2007ih}, to which we match with good accuracy, has been recently revisited in \cite{Amoretti:2020mkp}. The results in our case differ somewhat 
 from those in \cite{Amoretti:2020mkp}.  Notably they argue for the vanishing longitudinal DC conductivity; this corresponds to the limit $\tau\to\infty$ of \eqref{Hartnoll_sigmaxx}.} The underlying momentum sink of the adjoints makes hydrodynamic effective modeling untrustworthy at late times, unless the finite-$N_c$ corrections that are behind the long-time power law tails \cite{Kovtun:2003vj} are considered. To this vein, a less arduous approach would be to consider constructing a stationary state by forming an appropriate combination of a temperature gradient and an electric field \cite{Gouteraux:2018wfe}. This stationary state construction can be performed even in the translationally invariant system: the cancellation of the electric and heat currents causes zero net force and hence no momentum change. Then, the relativistic hydrodynamics description given in \cite{Hoyos:2020hmq} will allow one to find finite DC conductivities both in the quenched and in the fully unquenched case, corresponding to the 't Hooft limit $N_c\to\infty$ and Veneziano limit $N_f,N_c\to\infty,N_f/N_c$ fixed, respectively. This approach suggests that relativistic hydrodynamic models, perhaps coupled with scalar fields, should in principle be valid even at asymptotically large times.

We also wish to extend our search for striped phases in the gapped fractional quantum Hall (FQH) phases present in the current model when the internal fluxes are properly adjusted \cite{Bergman:2010gm}. A distinctive, novel property of the FQH phase is the presence of gapless neutral modes (but not  charged modes) in the bulk,\footnote{The existence of bulk neutral modes was found in a recent experiment \cite{Inoue}, in which highly sensitive noise measurement revealed the unexpected heat propagation through the incompressible FQH bulk at various filling factors.} in sharp contrast to all known QH or compressible states. A direct experimental observable consequence would then be that this electronic system is a bulk thermal conductor and a bulk charge insulator at the same time. This would interestingly contrast the known QH states that transport heat and charge along the edge but not in the bulk, and compressible states in which the bulk conducts both charge and heat.

More specifically, in an inhomogeneous FQH phase with orderly aligned stripes, we can ask whether the stripes slide in  presence of a temperature gradient and whether, in doing so, they support a heat current.  Furthermore, we could investigate whether the heat flows easier along the stripes than across them, as might be expected for a striped phase, or whether the heat transport stays surprisingly isotropic, as was observed for the charge transport in the gapless phase.   This avenue therefore adds motivation to compute the heat currents in the D3-D7' system, in addition to trying to understand the effectiveness of the hydrodynamic considerations discussed above.


\addcontentsline{toc}{section}{Acknowledgments}
\paragraph{Acknowledgments}

\noindent
We would like to thank Matteo Baggioli and Danny Brattan for discussions and helpful comments. N.~J.~is supported in part by the Academy of Finland grant no.~1322307. The work of M.~J.~was supported in part by an appointment to the JRG Program at the APCTP through the Science and Technology Promotion Fund and Lottery Fund of the Korean Government. M.~J.~was also supported by the Korean Local Governments -- Gyeongsangbuk-do Province and Pohang City, and by the National Research Foundation of Korea (NRF) funded by the Korean government (MSIT) (grant number 2021R1A2C1010834). M.~L.~is supported by National Science Foundation grant PHY-2000398.

\appendix

\section{Derivation of the DC conductivities}\label{app:DCderivation}

In this Appendix, we consider the DC conductivities for the striped background in the presence of a magnetic field. We concentrate on the unpinned case, in which the translation symmetry is not explicitly broken.   However, the results will also be useful for the pinned case, which essentially amounts to setting the sliding speed to zero, as we explain in the text. The derivation follows closely~\cite{Jokela:2016xuy}, where the expressions for the conductivities were derived at vanishing magnetic field.

We separate the explicit dependence on the magnetic field in the background by writing $a_y(x,u) = b x + \hat a_y(x,u)$. For the gauge field fluctuations we use the Ansatz~\cite{Araujo:2015hna,Jokela:2016xuy,Jokela:2017ltu}
\bea \label{Ansatz2}
 \delta a_y(t,x,u) &=& - E_y t +\delta a_y(x,u) - v_s t \partial_x \hat a_y(x,u) \\
\delta a_x(t,x,u) &=& -(E_x - p'(x))t + \delta a_x(x,u) \\ 
\label{eq:atfluct}
\delta a_t(t,x,u) & =& p(x) + \delta a_t(x,u) - v_s t\, \partial_x a_t(x,u)\\ 
\delta \psi(t,x,u) & =&  \delta \psi(x,u) - v_s t\, \partial_x \psi(x,u)\\ 
\delta z(t,x,u) & =&  \delta z(x,u) - v_s t\, \partial_x z(x,u)\ .\label{Ansatz3}
\eea
Here the fluctuations, the electric fields $E_x$ and $E_y$, and the function $p$ (which corresponds to the modulation of the chemical potential) are treated at linear order. The variation of the (averaged) chemical potential is set to zero, $\langle p \rangle =0$, where $\langle\,\cdots\rangle$ denotes average over $x$. Notice that, due to gauge covariance, $p$ does not appear in the fluctuation equations but it does affect the boundary conditions, as we shall see below.

As usual, the DC conductivities can be expressed in terms of horizon data by using conserved bulk currents. As it turns out, this works in a slightly different way for the (field theory) currents generated in $x$ and $y$ direction by the electric fields. We consider first the case of current in $x$ direction. 

It was shown in~\cite{Jokela:2016xuy} that, even in the presence of the sliding stripes (i.e., even at nonzero $v_s$), there is a quantity $\mathcal{J}_x$ which can be defined in terms on the fluctuations and is independent of both $x$ and $u$ on-shell. That is, the fluctuation equation for $a_x$, and the constraint corresponding to $a_u$, can be shown to imply
\be \label{eq:Jxeqs}
 \frac{\partial \mathcal{J}_x(x,u)}{\partial x} = 0 =  \frac{\partial \mathcal{J}_x(x,u)}{\partial u} \ ,
\ee
where
\be
 \mathcal{J}_x(x,u) = \frac{1}{\sqrt{2}}  \left(G_1(x,u)\ \partial_u \delta a_x(x,u)+v_s \tilde G_2(x,u) + E_y G_3(x,u)\right) 
\ee
so that $\mathcal{J}_x$ is constant when evaluated on the solutions to the fluctuation equations.
Here the coefficients $G_1$, $\tilde G_2$, and $G_3$ are functions of the background only. The expressions are somewhat unilluminating, but we include them here for completeness:
\begin{align}
 G_1 &= \frac{\sqrt{h \left(\frac{1}{2}+4 \sin^4\psi \right) \left(\frac{1}{2}+4 \cos^4\psi \right)}}{\sqrt{R}} h &\\
 \tilde G_2 & = -2 b c(\psi) +\frac{\sqrt{h \left(\frac{1}{2}+4 \sin^4\psi \right) \left(\frac{1}{2}+4 \cos^4\psi \right)}}{\sqrt{R}}\left[\partial_u a_t \left(1+b u^4 \partial_u  \hat a_y+b^2 u^4\right)-b u^4 \partial_x a_t \partial_u  \hat a_y\right]  & \\
 G_3 &= 2 c(\psi) -\frac{\sqrt{h \left(\frac{1}{2}+4 \sin^4\psi \right) \left(\frac{1}{2}+4 \cos^4\psi \right)}}{\sqrt{R}}u^4\left[ \partial_u a_t \left(\partial_x \hat a_y+b\right)-\partial_x a_t \partial_u \hat a_y\right] &
\end{align}
with
\begin{align}
 R & = -h \Big[u^4 (\partial_u a_t)^2 \left(2 b u^4 \hat a_y+u^4 \hat a_y^2+u^2 \psi ^2+z^2+b^2 u^4+1\right) & \nn\\
& -2 u^4 \partial_u a_t a_t \left(u^4 \partial_u \hat a_y \left(\partial_x \hat a_y+b\right)+u^2 \partial_u \psi  \partial_x \psi +\partial_u z \partial_x z\right)+u^8 (\partial_x a_t)^2 (\partial_u \hat a_y)^2& \nn\\
&+u^6 (\partial_x a_t)^2 (\partial_u \psi )^2+u^4 (\partial_x a_t)^2 (\partial_u z)^2-2 b u^4 \partial_x \hat a_y-u^4 (\partial_x \hat a_y)^2-u^2 (\partial_x \psi )^2-z^2-b^2 u^4-1\Big] & \nn\\
& -u^4 (\partial_x a_t)^2+h^2 \Big[u^2 (\partial_u \psi )^2 \left(2 b u^4 \partial_x \hat a_y+u^4 (\partial_x \hat a_y)^2+(\partial_x z)^2+b^2 u^4+1\right)& \nn\\
&+(\partial_u z)^2 \left(2 b u^4 \partial_x \hat a_y+u^4 (\partial_x  \hat a_y)^2+u^2 (\partial_x \psi )^2+b^2 u^4+1\right)& \nn\\
&-2 u^4 \partial_u \hat a_y \left(\partial_x \hat a_y+b\right) \left(u^2 \partial_u \psi  \partial_x \psi +\partial_u z \partial_x z\right)+u^4 (\partial_u \hat a_y)^2 \left(u^2 (\partial_x \psi )^2+(\partial_x z)^2+1\right)& \nn\\
&-2 u^2 \partial_u \psi  \partial_x \psi  \partial_u z \partial_x z\Big]\ . &
\end{align}

The boundary value of the conserved bulk charge evaluates to
\be \label{conservcondbdry}
 \lim_{u\to 0} \mathcal{J}_x(x,u)  = j_x(x) - v_s d(x) \ ,
\ee
where $j_x(x) = \partial_u \delta a_x(x,0)$ is the electric current, and the second term appears due to the sliding of the (inhomogeneous) background charge density $d(x) = -\partial_u a_t(x,0)$. Because $\mathcal{J}_x$ is independent of $u$ when evaluated on the solution, the modulation of the current is therefore only given by the movement of the charge density in the stripes. 

At the horizon we define
\be \label{deltaaxhor}
 \delta a_x(x,u) = \delta a_{x,0}(x) \log(1-u)+\morder{(1-u)^0}\ .
\ee
for the fluctuations and
\begin{align} \label{bghorexp1}
 \psi(x,u) &= \psi_0(x) + \morder{1-u}\ ,& \qquad  z(x,u) &= z_0(x) + \morder{1-u}\ , & \\
  a_y(x,u) &= b x +a_{y,0}(x) + \morder{1-u}\ ,&\qquad a_t(x,u) &= a_{t,0}(x)(u-1) +\morder{(1-u)^2}
 \label{bghorexp2}
\end{align}
for the background. 
We find that 
\bea
 &&\lim_{u\to 1} \mathcal{J}_x(x,u) = \sqrt{2} \left(E_y -v_s b\right) c(\psi_0(x)) \nn\\
 &&\qquad \qquad-  \left( 4 \delta a_{x,0}(x) + \left(E_y -v_s b\right) a_{t,0}(x) (b+a_{y,0}'(x)) -v_s a_{t,0}(x) \right)\hat \sigma(x) \ ,
\eea
where 
\be 
 \hat\sigma(x) = \frac{ \sqrt{\left(1+8 \sin ^4 \psi_0(x)\right) \left(1+8 \cos ^4 \psi_0(x)\right) }}{2\sqrt{2\left(1- a_{t,0}(x)^2\right) \left(1+(b+a_{y,0}'(x))^2+\psi_0'(x)^2+z_0'(x)^2\right) }} \ .  
 \label{localcond1}
 \ee
Notice that the combination $E_y -v_s b$ is the (Galilean) boosted electric field.
Further IR regularity implies that (see~\cite{Jokela:2016xuy}) 
\be \label{regcond}
 \delta a_{x,0}(x) = -\frac 14 \left(E_x - p'(x)\right)\ .
\ee

Equating the UV and IR values of the conserved quantity $\mathcal{J}_x$ gives
\begin{align}
 j_c &= j_x(x) - v_s d(x)&  \\
   &= \sqrt{2} c(\psi_0(x)) \left(E_y -v_s b\right)  &  \nonumber\\
   &\phantom{=}\, + \left(E_x -p'(x)+ v_s a_{t,0}(x)- \left(E_y -v_s b\right) a_{t,0}(x) (b+a_{y,0}'(x)) \right)\hat \sigma(x) \ ,
\end{align}
where $j_c$ denotes the constant value of the current.
We can eliminate $p$ by taking the average over $x$, after dividing first by $\hat \sigma$. This leads to 
 \be\label{jccond}%
 j_c\langle \hat\sigma^{-1}\rangle = E_x+v_s\langle a_{t,0}\rangle +\left[\sqrt{2} \langle c(\psi_0) \hat \sigma^{-1} \rangle - \langle a_{t,0} (b+a_{y,0}') \rangle \right] \left(E_y -v_s b\right)\ ,
 \ee

The generic expression for $p'(x)$ becomes
\begin{align} \label{pprimegen}
 p'(x) &= E_x\left(1-\frac{\langle\hat \sigma^{-1} \rangle^{-1}}{\hat \sigma(x)}\right) + v_s\left( a_{t,0}(x) - \langle a_{t,0}\rangle \frac{\langle\hat \sigma^{-1} \rangle^{-1}}{\hat \sigma(x)}\right)+\left(E_y -v_s b\right)\Bigg(\frac{\sqrt{2}c(\psi_0(x))}{\hat \sigma(x)} &\nn\\ 
 & -\frac{\sqrt{2}\langle c(\psi_0) \hat \sigma^{-1} \rangle\langle\hat \sigma^{-1} \rangle^{-1}}{\hat \sigma(x)} - a_{t,0}(x) (b+a_{y,0}'(x)) + \langle a_{t,0} (b+a_{y,0}') \rangle\frac{\langle\hat \sigma^{-1} \rangle^{-1}}{\hat \sigma(x)}\Bigg)  \,.&
\end{align}
After eliminating $p$, the current $j_x(x)$ is solved as
\begin{align}
 j_x(x) &= \left(d(x) + \langle a_{t,0}\rangle \langle\hat \sigma^{-1} \rangle^{-1}\right)v_s + \langle\hat \sigma^{-1} \rangle^{-1} E_x &\nn\\
  &\phantom{=} + \left[\sqrt{2} \langle c(\psi_0) \hat \sigma^{-1} \rangle - \langle a_{t,0} (b+a_{y,0}') \rangle \right]\langle\hat \sigma^{-1} \rangle^{-1}  \left(E_y -v_s b\right) \ . &
\end{align}
The current cannot be expressed solely in terms of horizon quantities, but as for $b=0$, we can use the relation (see~\cite{Jokela:2016xuy})
\be \label{chargeidentity}
\langle d \rangle 
= \sqrt{2} \langle c(\psi_0) (b+a_{y,0}') \rangle -  \langle a_{t,0} \hat \sigma \left(1+(b+a_{y,0}')^2+\psi_0'^2+z_0'^2\right)\rangle
\ee
to write it in a form where the only boundary quantity is the modulation of charge density:
\begin{align} \label{jxfinal}
 j_x(x) &= \left(d(x) -\langle d \rangle\right)v_s +\bigg[\sqrt{2} \langle c(\psi_0) (b+a_{y,0}') \rangle  & \nn \\
  &\phantom{=}-  \langle a_{t,0} \hat \sigma \left(1+(b+a_{y,0}')^2+\psi_0'^2+z_0'^2\right)\rangle + \langle a_{t,0}\rangle \langle\hat \sigma^{-1} \rangle^{-1}\bigg]v_s & \nn \\
  &\phantom{=}+ \langle\hat \sigma^{-1} \rangle^{-1} E_x 
   + \left[\sqrt{2} \langle c(\psi_0) \hat \sigma^{-1} \rangle - \langle a_{t,0} (b+a_{y,0}') \rangle \right]\langle\hat \sigma^{-1} \rangle^{-1}  \left(E_y -v_s b\right) \ . &
\end{align}
The separation of the charge to two terms in~\eqref{chargeidentity} suggests that the first term should be interpreted as the induced charge and the second as the ordinary charge.

We then analyze the current $j_y$. One can define a quantity analogous to $\mathcal{J}_x$ but related to $a_y$ rather than $a_x$:
\be
\mathcal{J}_y (x,u) = \frac{1}{\sqrt{2}}\frac{\delta S}{\delta ( \partial_u \delta a_y(t,x,u) )}  \ .
\ee
However, unlike $\mathcal{J}_x$, this quantity does not take a constant value everywhere, which means that we cannot relate the current $j_y$ to horizon quantities pointwise in $x$.
Instead, because the action depends on $\delta a_y$ only through its derivatives, the fluctuation equation for $\delta a_y$ takes the form 
\be
\sqrt{2}\frac{\partial}{\partial u} \mathcal{J}_y (x,u) + \frac{\partial}{\partial x} \frac{\delta S}{\delta ( \partial_x \delta a_y(t,x,u) )} = 0 \ ,
\ee
where the second term is periodic in $x$. That is, there is a bulk current with two components that is conserved. Therefore, integrating this equation over $x$, we find that the spatially averaged charge
$\left\langle\mathcal{J}_y (x,u) \right\rangle$ is conserved; i.e., it is independent of $u$ when evaluated on the solutions to the fluctuation equations.
Consequently, we can compute averaged Hall currents using a similar procedure as used above for the longitudinal currents.
We do not include the explicit expression for $\mathcal{J}_y$ because it is very complicated. It can be expressed as a linear combination of $E_x$ and (derivatives of) the various fluctuations, with coefficients depending on the background.

The UV limit is given by 
\be \label{jyuto0}
 \lim_{u \to 0}\mathcal{J}_y (x,u)  = \partial_u \delta a_y(x,0)\equiv  \bar j_y(x) =  j_y(x,t) + v_s t J_y'(x) \,,
\ee
where $\bar j_y(x)$ is the current due to the time-independent fluctuation  $\delta a_y(x,u)$, $j_y(x,t)$ is the full fluctuated current corresponding to~\eqref{eq:atfluct}, and  $J_y(x) = \partial_u a_y(x,0)$ is the modulated current of the background. 
The time-dependent background current term reflects the delta peaks of the optical conductivities at zero frequency. As pointed out in~\cite{Jokela:2016xuy}, the current $j_y$ is actually ambiguous because of the freedom in the choice of the origin of the time-dependent term (we used above $t=0$). The ambiguity however disappears when averaging over $x$, since $\langle J_y'\rangle = 0$, so that also $\langle \bar j_y \rangle = \langle j_y \rangle$. That is, we are able to compute only averaged conductivities for current in the $y$ direction, but actually only the average is well-defined.

In order to compute the IR limit, we need the IR expansions for all fluctuations (expect for $\delta a_x$). We write\footnote{Notice that the term $-p(x)$ here cancels a similar term in~\eqref{Ansatz3} so that the full fluctuation $\delta a_{t}(t,x,u)$ vanishes at the horizon, which is expected because the background field $a_t$ also vanishes.}
\be
 \delta a_{t}(x,u) = -p(x) + \morder{1-u}
\ee
and
\be
 \delta f(x,u) = \delta f_0(x) \log(1-u) + \morder{(1-u)^0}
\ee
for the other fields. Then the IR limit becomes
\bea \label{jyuto1}
 &&\lim_{u \to 1} 
 \mathcal{J}_y(x,u)
 = -\sqrt{2} c(\psi_0(x))\left(E_x - p'(x)\right) \\\nn
&& + \hat \sigma(x) \big[ - 4 (1+z_0'(x)^2+\psi_0'(x)^2) \delta a_{y,0}(x) + (E_x-p'(x))a_{t,0}(x) (b+a_{y,0}'(x)) \\\nn
&& + 4 (b+a_{y,0}'(x))( z_0'(x) \delta z_0(x)+  \psi_0'(x) \delta \psi_0(x))\big] \ .
\eea
The regularity conditions at the horizon, which generalize~\eqref{regcond}, arise from the terms proportional to $t$ in our Ansatz,
\be
 \delta a_{y,0}(x) = -\frac{E_y}{4} - \frac{v_s}{4} a_{y,0}'(x) \,,\quad \delta \psi_0(x) = -\frac{v_s}{4} \psi_0'(x)\,,\quad \delta z_0(x) = -\frac{v_s}{4} z_0'(x) \ .
\ee
Inserting these and $p'(x)$ from~\eqref{pprimegen} in~\eqref{jyuto1}, taking the average over $x$, and equating with the UV limit~\eqref{jyuto0} we find
\begin{align} \label{jyfinal}
 \langle j_y\rangle 
 &= \Bigg[\left\langle \hat \sigma(1+z_0'^2+\psi_0'^2) + \frac{1}{\hat \sigma}\left(\sqrt{2}c(\psi_0) 
 - \hat \sigma a_{t,0} (b+a_{y,0}'(x))\right)^2 \right\rangle  & \nn \\
 &\phantom{=}-  \langle\hat \sigma^{-1} \rangle^{-1} \left(\sqrt{2} \langle c(\psi_0) \hat \sigma^{-1} \rangle - \langle a_{t,0} (b+a_{y,0}') \rangle \right)^2 \Bigg]  \left(E_y -v_s b\right) &\nn\\
 & \phantom{=} - \left[\sqrt{2} \langle c(\psi_0) \hat \sigma^{-1} \rangle - \langle a_{t,0} (b+a_{y,0}') \rangle \right]\langle\hat \sigma^{-1} \rangle^{-1} E_x  & \nn\\
 &\phantom{=} + \Bigg[ 
 \left(\langle a_{t,0} (b+a_{y,0}') \rangle-\sqrt{2} \langle c(\psi_0) \hat \sigma^{-1} \rangle\right)\langle a_{t,0} \rangle\langle\hat \sigma^{-1} \rangle^{-1} &\nn \\
  &\phantom{=}  + \sqrt{2} \langle a_{t,0}c(\psi_0)\rangle + \langle\hat\sigma(b+a_{y,0}') \left(1-a_{t,0}^2 
  \right)\rangle \Bigg]v_s
\end{align}
 
When parity is broken by the magnetic field, the speed $v_s$ will be induced both by $E_x$ and by $E_y$. That is, we write
\be
\label{slidingspeedapp}
 v_s = \hat v_x E_x + \hat v_y E_y
\ee
where the coefficients $\hat v_i$ are independent of the electric field (at linear order). In terms of these coefficients, the final expressions for the DC conductivities read:
\begingroup
\allowdisplaybreaks
\begin{align}
 \label{finalDCsigmaxx}
 \sigma_{xx}^\mathrm{DC}(x) &= \left(d(x) -\langle d \rangle\right) \hat v_x +\bigg[\sqrt{2} \langle c(\psi_0) (b+a_{y,0}') \rangle  & \nn \\
  &\phantom{=}-  \langle a_{t,0} \hat \sigma \left(1+(b+a_{y,0}')^2+\psi_0'^2+z_0'^2\right)\rangle + \langle a_{t,0}\rangle \langle\hat \sigma^{-1} \rangle^{-1}\bigg]\hat v_x & \nn \\
  &\phantom{=}+ \langle\hat \sigma^{-1} \rangle^{-1}  
   - \left[\sqrt{2} \langle c(\psi_0) \hat \sigma^{-1} \rangle - \langle a_{t,0} (b+a_{y,0}') \rangle \right]\langle\hat \sigma^{-1} \rangle^{-1} \hat v_x b  &\\
\label{finalDCsigmaxy}
    \sigma_{xy}^\mathrm{DC}(x) &= \left(d(x) -\langle d \rangle\right)\hat v_y+\bigg[\sqrt{2} \langle c(\psi_0) (b+a_{y,0}') \rangle  & \nn \\
  &\phantom{=}-  \langle a_{t,0} \hat \sigma \left(1+(b+a_{y,0}')^2+\psi_0'^2+z_0'^2\right)\rangle + \langle a_{t,0}\rangle \langle\hat \sigma^{-1} \rangle^{-1}\bigg]\hat v_y & \nn \\
  &\phantom{=} 
   + \left[\sqrt{2} \langle c(\psi_0) \hat \sigma^{-1} \rangle - \langle a_{t,0} (b+a_{y,0}') \rangle \right]\langle\hat \sigma^{-1} \rangle^{-1}  \left(1 -\hat v_y b\right)  &\\
\label{finalDCsigmayy}
\langle\sigma_{yy}^\mathrm{DC}\rangle &= \Bigg[\left\langle \hat \sigma(1+z_0'^2+\psi_0'^2) + \frac{1}{\hat \sigma}\left(\sqrt{2}c(\psi_0) 
 - \hat \sigma a_{t,0} (b+a_{y,0}')\right)^2 \right\rangle  & \nn \\
 &\phantom{=}-  \langle\hat \sigma^{-1} \rangle^{-1} \left(\sqrt{2} \langle c(\psi_0) \hat \sigma^{-1} \rangle - \langle a_{t,0} (b+a_{y,0}') \rangle \right)^2 \Bigg]  \left(1 -\hat v_y b\right) &\nn\\
 &\phantom{=} + \Bigg[ 
 \left(\langle a_{t,0} (b+a_{y,0}') \rangle-\sqrt{2} \langle c(\psi_0) \hat \sigma^{-1} \rangle\right)\langle a_{t,0} \rangle\langle\hat \sigma^{-1} \rangle^{-1} &\nn \\
  &\phantom{=}  + \sqrt{2} \langle a_{t,0}c(\psi_0)\rangle + \langle\hat\sigma(b+a_{y,0}') \left(1-a_{t,0}^2 
  \right)\rangle \Bigg]\hat v_y &\\
  \label{finalDCsigmayx}
  \langle\sigma_{yx}^\mathrm{DC}\rangle &= -\Bigg[\left\langle \hat \sigma(1+z_0'^2+\psi_0'^2) + \frac{1}{\hat \sigma}\left(\sqrt{2}c(\psi_0) 
 - \hat \sigma a_{t,0} (b+a_{y,0}')\right)^2 \right\rangle  & \nn \\
 &\phantom{=}-  \langle\hat \sigma^{-1} \rangle^{-1} \left(\sqrt{2} \langle c(\psi_0) \hat \sigma^{-1} \rangle - \langle a_{t,0} (b+a_{y,0}') \rangle \right)^2 \Bigg]   \hat v_x b &\nn\\
 & \phantom{=} - \left[\sqrt{2} \langle c(\psi_0) \hat \sigma^{-1} \rangle - \langle a_{t,0} (b+a_{y,0}') \rangle \right]\langle\hat \sigma^{-1} \rangle^{-1}   & \nn\\
 &\phantom{=} + \Bigg[ 
 \left(\langle a_{t,0} (b+a_{y,0}') \rangle-\sqrt{2} \langle c(\psi_0) \hat \sigma^{-1} \rangle\right)\langle a_{t,0} \rangle\langle\hat \sigma^{-1} \rangle^{-1} &\nn \\
  &\phantom{=}  + \sqrt{2} \langle a_{t,0}c(\psi_0)\rangle + \langle\hat\sigma(b+a_{y,0}') \left(1-a_{t,0}^2 
  \right)\rangle \Bigg] \hat v_x\ .
\end{align}
\endgroup
Notice that for the first two conductivities \eqref{finalDCsigmaxx} and  \eqref{finalDCsigmaxy}, taking the spatial average simply eliminates the first terms proportional to $d(x) -\langle d \rangle$.

\bibliographystyle{JHEP}
\bibliography{refs}

\end{document}